\documentclass[
    reprint,
    twocolumn,
    superscriptaddress,
    amsmath, amssymb,
    aps,
    prx,
    floatfix,
]{revtex4-2}

\usepackage{newtxtext, newtxmath}
\usepackage[usenames, dvipsnames]{xcolor}
\usepackage{graphicx}
\usepackage{dcolumn} 
\usepackage{xpatch}
    \makeatletter \xpretocmd \start@align{\linenomathWithnumbers}{}{\fail}
\usepackage{mathtools, url}
\usepackage{CJKutf8}
\usepackage[toc]{appendix}
\usepackage[ruled, linesnumbered]{algorithm2e}
\usepackage[unicode=true, bookmarks=true, bookmarksnumbered=false, bookmarksopen=false, breaklinks=false, pdfborder={0 0 1}, backref=false, colorlinks=true]{hyperref}
    \hypersetup{linkcolor=black, urlcolor=black, citecolor=black, pdfstartview={FitH}, unicode=true}

\usepackage[math]{cellspace}
\renewcommand{\vec}[1]{\boldsymbol{#1}}
\renewcommand{\tensor}[1]{\mathbf{#1}}
\newcommand{\DD}{\mathrm{D}}
\newcommand{\p}{\partial}
\newcommand{\dif}{\mathop{}\!\mathrm{d}}
\newcommand{\bn}{\vec{\nabla}}
\renewcommand{\widebar}[1]{\mskip.5\thinmuskip\overline{\mskip-.5\thinmuskip {#1} \mskip-.5\thinmuskip}\mskip.5\thinmuskip}
\newcommand{\ee}{\mathrm{e}}
\newcommand{\ii}{\mathrm{i}}
\newcommand{\ave}[1]{\langle #1 \rangle}

\newcommand{\EQ}{\begin{equation}}
\newcommand{\EN}{\end{equation}}

\renewcommand{\le}{\leqslant}
\newcommand{\ket}[1]{| #1 \rangle}

\newcommand{\T}{^{\mathrm{T}}}
\newcommand{\dx}{\Delta x}

\newcommand{\dt}{\Delta t}
\newcommand{\norm}[1]{\| #1 \|}

\allowdisplaybreaks

\begin{document}

\title{Quantum spin representation for the Navier-Stokes equation}

\author{Zhaoyuan Meng}
    \affiliation{State Key Laboratory for Turbulence and Complex Systems, College of Engineering, Peking University, Beijing 100871, PR China}
\author{Yue Yang}
    \email{yyg@pku.edu.cn}
    \affiliation{State Key Laboratory for Turbulence and Complex Systems, College of Engineering, Peking University, Beijing 100871, PR China}
    \affiliation{HEDPS-CAPT, Peking University, Beijing 100871, PR China}

\date{\today}

\begin{abstract}
We develop a quantum representation for Newtonian viscous fluid flows by establishing a mapping between the Navier-Stokes equation (NSE) and the Schr\"odinger-Pauli equation (SPE).
The proposed nonlinear SPE incorporates the two-component wave function and the imaginary diffusion.
Consequently, classical fluid flow can be interpreted as a non-Hermitian quantum spin system.
Using the SPE-based numerical simulation of viscous flows, we demonstrate the quantum/wave-like behavior in flow dynamics.
Furthermore, the SPE equivalent to the NSE can facilitate the quantum simulation of fluid dynamics.
\end{abstract}

\maketitle

\section{Introduction}
Fluid dynamics and quantum mechanics represent physics at macroscopic and microscopic scales, respectively, but they can demonstrate remarkable similarities.
Some macroscopic flows exhibit quantum phenomena, such as the wave-particle duality exhibited by walking droplets~\cite{Gilet2009_Chaotic, Bush2020_Hydrodynamic, Saenz2021_Emergent}, the Casimir effect observed in turbulence and turbulent plasmas~\cite{Kardar1999_The, Davoodianidalik2022_Fluctuation, Mendonca2001_Casimir}, and the Saffman-Taylor fingering akin to the growth of an electronic droplet in the quantum Hall regime~\cite{Agam2002_Viscous}.
Conversely, quantum spin systems can display emergent hydrodynamic behavior~\cite{Zu2021_Emergent, Joshi2022_Observing}. The Schrödinger-like quantum representation has proven useful in capturing certain flow phenomena, including superfluids~\cite{Gross1961_Structure, Pitaevskii1961_Vortex}, quantum plasmas~\cite{Marklund2007_Dynamics}, non-Abelian fluids~\cite{Love2004_Quaternionic}, soliton dynamics of vortex filaments~\cite{Hasimoto1972_A}, and topological active matter~\cite{Sone2019_Anomalous}.

In mathematics, fluid dynamics and quantum mechanics are described by two seemingly distinct equations, the Navier-Stokes equation (NSE) and the Schrödinger equation, respectively.
Despite their apparent differences, some attempts have been made to establish an explicit mathematical connection between the two fields.
Madelung~\cite{Madelung1927_Quantentheorie} recast the Schrödinger equation into a modified Euler equation for compressible potential flows, but this transform has little application due to the lack of the vorticity, a key component in shear flows and turbulence.

Several works have striven to incorporate the finite vorticity and viscous effects into the quantum representation of fluid dynamics.
Sorokin~\cite{Sorokin2001_Madelung} extended the Madelung transform to account for finite vorticity using the two-component wave function.
Chern \emph{et al.}~\cite{Chern2016_Schrodinger, Chern2017_Inside} employed the two-component wave function to characterize the incompressible Schrödinger flow with an artificial body force.
Meng and Yang~\cite{Meng2023_Quantum} generalized the incompressible Schrödinger flow to compressible flows.
Salasnich \emph{et al.}~\cite{Salasnich2024_Quantum} introduced the imaginary diffusion of the wave function to achieve the correct dissipation for Newtonian fluids, but they only examined the single-component wave function for potential flows.

\begin{figure*}
    \centering
    \includegraphics{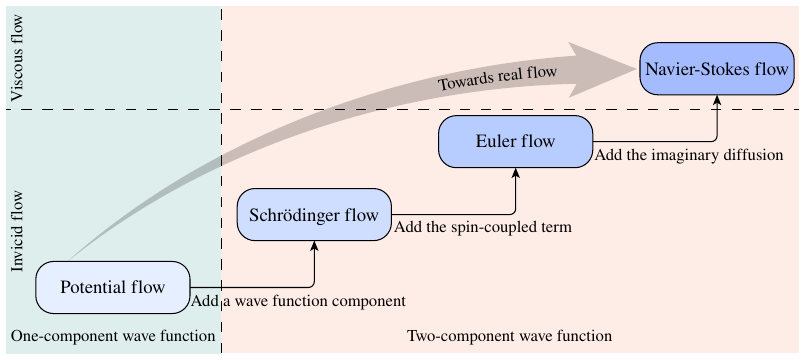}
    \caption{Progression of the quantum representation for fluid dynamics, from simple to complex flows.}
    \label{fig:qm-fm_supp}
\end{figure*}

Here we develop a comprehensive quantum representation for real fluid flows by establishing an explicit mapping between the NSE and the Schr\"odinger-Pauli equation (SPE) with the two-component wave function.
The proposed nonlinear SPE for a quantum spin system is a Hamiltonian formulation~\cite{Morrison1998_Hamiltonian} for both compressible and incompressible viscous flow with finite vorticity.
The imaginary diffusion~\cite{Salasnich2024_Quantum} is incorporated into the SPE, leading to a non-Hermitian quantum system.
The gap between the quantum representations of the Schrödinger flow~\cite{Chern2016_Schrodinger, Chern2017_Inside, Meng2023_Quantum} and the real flow is resolved in this paper.
The progression of the quantum representation for fluid dynamics, from simple to complex flows, is sketched in Fig.~\ref{fig:qm-fm_supp} and will be elaborated below.

The classical Madelung transform~\cite{Madelung1927_Quantentheorie} is related to the de Broglie-Bohm interpretation of quantum mechanics~\cite{Bohm1952_A, Takabayasi1952_On, Takabayasi1953_Remarks, Wyatt2005_Quantum, Bohm2012_Quantum}.
Conversely, the present NSE-SPE mapping can facilitate quantum computing of fluid dynamics through Hamiltonian simulation~\cite{Meng2023_Quantum, Lu2023_Quantum}, in particular, on the current noisy intermediate-scale quantum (NISQ) hardware \cite{Meng2024_Simulating}.
%
%
Thus, the quantum simulation~\cite{Feynman1982_Simulating} of fluid flows (as a unique quantum spin system) holds immense potential in various applications, e.g., in aerospace engineering and weather forecasting~\cite{Givi2020_Quantum, Succi2023_Quantum, Bharadwaj2023_Hybrid}. 
Moreover, simplified forms of the NSE-SPE mapping have been used in investigating the potential finite-time singularity in the Euler equation~\cite{Meng2024_Lagrangian} and developing fast algorithms for fluid simulation and animation~\cite{Chern2016_Schrodinger, Chern2017_Inside, Xiong2022Clebsch}.

The outline of the present paper is as follows.
Section~\ref{sec:SPE-NSE} introduces the Schr\"odinger-Pauli formulation of the NSE.
Section~\ref{sec:SPE_evolution} illustrates the quantum behaviour in the flow evolution using the SPE-based numerical simulation.
Some discussions are presented in Sec.~\ref{sec:discussion}.

\section{Hydrodynamic formulation of the SPE}\label{sec:SPE-NSE}
\subsection{SPE for compressible flow}
We employ the hydrodynamic formulation of the SPE~\cite{Pesci2005_Mapping, Meng2023_Quantum} to represent the compressible/incompressible NSE with finite vorticity, highlighting the essential role of the vorticity $\vec{\omega}\equiv\bn\times\vec{u}$ in viscous flows~\cite{Saffman1993_Vortex}, with the fluid velocity $\vec{u}$.
The quantum system governed by the SPE is characterized by a two-component spinor wave function $\ket{\psi} \equiv [\psi_+(\vec{x},t), \psi_-(\vec{x},t)]\T$, where $\psi_+$ and $\psi_-$ are the spin-up and -down components, respectively~\cite{Mueller2002_Two, Kasamatsu2003_Vortex, Liu2019_Schrodinger}.
The probability current is given by $\vec{J}\equiv \mathrm{Re}\langle \psi|\hat{\vec{p}}|\psi\rangle / m$, with the particle mass $m$, the momentum operator in the position representation $\hat{\vec{p}}=-\ii\hbar\bn$, and the reduced Planck constant $\hbar$.
The flow velocity is then defined as
\begin{equation}\label{eq:vel}
    \vec{u} \equiv \frac{\vec{J}}{\rho}
    = \frac{1}{m}\frac{\mathrm{Re}\langle \psi|\hat{\vec{p}}|\psi\rangle}{\langle\psi|\psi\rangle}
\end{equation}
via the generalized Madelung transform~\cite{Meng2023_Quantum}, where
\begin{equation}\label{eq:rho}
    \rho
    \equiv \langle\psi|\psi\rangle
    = |\psi_+|^2+|\psi_-|^2
\end{equation}
is the fluid density.

We recast the compressible (i.e., general) NSE
\begin{equation}\label{eq:compressible_NS}
    \begin{dcases}
        \frac{\p\rho}{\p t} + \bn\cdot(\rho\vec{u}) = 0, \\
        \frac{\p\vec{u}}{\p t} + \vec{u}\cdot\bn\vec{u} = -\frac{1}{\rho}\bn p + \frac{\mu}{3\rho}\bn(\bn\cdot\vec{u}) + \frac{\mu}{\rho}\Delta\vec{u}
    \end{dcases}
\end{equation}
into a nonlinear SPE
\begin{equation}\label{eq:Schrodinger_2}
    \begin{aligned}
        \ii\hbar\frac{\p}{\p t}\ket{\psi}
        =\ & \bigg[ \bigg(1-\ii\frac{2m\mu}{\hbar\rho}\bigg)\frac{\hat{\vec{p}}^2}{2m}
        + \tilde{p} + \frac{\hbar^2}{2m\rho}|\bn\psi|^2
        \\
        &+ \ii\frac{\hbar\mu}{2}\frac{2|\bn\psi|^2 - \Delta\rho}{\rho^2}\bigg] \tensor{I}\ket{\psi}
        + \vec{\sigma}\cdot\vec{B}\ket{\psi}.
    \end{aligned}
\end{equation}
The derivation for this hydrodynamic SPE is detailed in Appendix~\ref{app:derivation_SPE_NSE}.
Here, $\mu$ is a constant viscosity coefficient,
$|\bn\psi|^2=|\bn\psi_+|^2+|\bn\psi_-|^2$ is the 2-norm of the gradient of $\ket{\psi}$,
$\tilde{p} = P - m|\vec{u}|^2/2 + \hbar\vec{s}\cdot\vec{f}/\rho$  is a modified pressure
with $P=\int m/\rho\dif\left[p - \hbar^2|\bn\rho|^2/(4m^2\rho) - \mu\bn\cdot\vec{u}/3 \right]$, the static pressure $p$ given by the equation of state, and a vector $\vec{f}\in\mathbb{R}^3$ to be determined,
$\tensor{I}$ is the $2\times 2$ identity matrix,
$\vec{\sigma}=\sigma^j\vec{e}_j$ is the Pauli vector,
$\vec{B}=\hbar^2\Delta\vec{s}/(4m\rho) - \hbar\vec{f}$ is a spin-coupled vector, and
\begin{equation}\label{eq:spin_vector}
    \vec{s} \equiv \langle\psi|\vec{\sigma}|\psi\rangle
\end{equation}
is the spin vector (or the average spin).
Note that Eq.~\eqref{eq:Schrodinger_2} is closed with the equations of state and energy in general, and when $p$ is solely determined by $\psi$ in particular, such as in the quantum potential~\cite{Wyatt2005_Quantum, Bohm2012_Quantum}.

We employ the Moore-Penrose solution for $\vec{f}$ as
\begin{equation}\label{eq:f}
    \vec{f} = \tensor{A}^+\cdot\vec{\mathcal{S}},
\end{equation}
which avoids the singularity of $|\tensor{A}|=0$ and minimizes $\|\vec{f}\|_2$ for the numerical stability, with matrix elements $A_{ij}=\p_is_j - s_j\p_i\ln\rho$, the Moore-Penrose pseudoinverse $\tensor{A}^+$, and
\begin{equation*}
    \begin{aligned}
        \vec{\mathcal{S}}
        =\ & \frac{\hbar}{4m\rho^2}\bn\rho\cdot\bigg[\bn(\rho\bn\rho) - \bn\bn\vec{s}\cdot\vec{s} + \frac{8m^2\mu}{\hbar^2}\rho\bn\vec{u} \bigg]
        \\
        &+ \frac{2m\mu}{\hbar\rho}|\bn\psi|^2\vec{u}
        - \frac{m\mu}{\hbar}(\bn\cdot\vec{u} + \vec{u}\cdot\bn\ln\rho)\bn\ln\rho.
    \end{aligned}
\end{equation*}

The energy expectation of the quantum system in Eq.~\eqref{eq:Schrodinger_2} is
\begin{equation}
    \begin{aligned}
        \ave{E} =\ & \int_{\mathcal{D}}\bigg[\frac{m\rho|\vec{u}|^2}{2} + \rho P + \frac{\hbar^2}{4m}\bigg(\Delta\rho + \frac{|\bn\rho|^2}{\rho} \bigg)\bigg] \dif\vec{x}
        \\
        &+ \ii\hbar\mu\int_{\mathcal{D}}\vec{u}\cdot\bn\ln\rho \dif\vec{x}
    \end{aligned}
\end{equation}
in the solution domain $\mathcal{D}$.
Hence, the Hamiltonian of Eq.~\eqref{eq:Schrodinger_2} is non-Hermitian for $\mu\ne 0$, as $\ave{E}$ is complex-valued.
The real part of $\ave{E}$ comprises the total kinetic energy, the pressure energy, and the compressible effect.

For the inviscid flow with $\mu=0$, Eq.~\eqref{eq:compressible_NS} degenerates to the compressible Euler equations
\begin{equation}
    \begin{dcases}
        \frac{\p\rho}{\p t} + \bn\cdot(\rho\vec{u}) = 0, \\
        \frac{\p\vec{u}}{\p t} + \vec{u}\cdot\bn\vec{u} = -\frac{1}{\rho}\bn p,
    \end{dcases}
\end{equation}
and the corresponding SPE reduces to
\begin{equation}
    \ii\hbar\frac{\p}{\p t}\ket{\psi}
    = \bigg( \frac{\hat{\vec{p}}^2}{2m}
    + \tilde{p} + \frac{\hbar^2}{2m\rho}|\bn\psi|^2 \bigg)\tensor{I}\ket{\psi}
    + \vec{\sigma}\cdot\vec{B}\ket{\psi},
\end{equation}
with the modified pressure
\begin{equation}\label{eq:SM_tildep_compressible_inviscid}
    \tilde{p} = \int\frac{m}{\rho}\dif\left(p - \frac{\hbar^2}{4m^2\rho}|\bn\rho|^2 \right) - \frac{m}{2}|\vec{u}|^2 + \frac{\hbar}{\rho}\vec{s}\cdot\vec{f},
\end{equation}
and $\vec{f}=\tensor{A}^+\cdot\vec{\mathcal{S}}$ with $\tensor{A}$ in Eq.~\eqref{eq:SM_A_compressible} and
\begin{equation}\label{eq:SM_compressible_inviscid_S}
    \vec{\mathcal{S}} = \frac{\hbar}{4m\rho^2}\bn\rho\cdot\left[\bn(\rho\bn\rho) - \bn\bn\vec{s}\cdot\vec{s} \right].
\end{equation}

\subsection{SPE for incompressible flow}
A particular case of Eq.~\eqref{eq:compressible_NS} is for the incompressible flow with constant density $\rho=\rho_0$.
Here, $\ket{\psi}$ resembling the momentum eigenstate (i.e., the plane wave) is subjected to a solenoidal constraint $\mathrm{Im}\langle\psi|\hat{\vec{p}}^2|\psi\rangle = 0$ for the divergence-free velocity.
The incompressible NSE
\begin{equation}\label{eq:incompressible_NS}
    \begin{dcases}
        \bn\cdot\vec{u} = 0, \\
        \frac{\p\vec{u}}{\p t} + \vec{u}\cdot\bn\vec{u} = -\bn \frac{p}{\rho_0} + \nu\Delta\vec{u}
    \end{dcases}
\end{equation}
with constant kinematic viscosity $\nu\equiv \mu/\rho_0$ is reformulated into a nonlinear SPE
\begin{equation}\label{eq:Schrodinger_4}
    \begin{aligned}
        \ii\hbar\frac{\p}{\p t}\ket{\psi}
        =\ & \bigg[\bigg(1 - \ii\frac{2m\nu}{\hbar} \bigg)\frac{\hat{\vec{p}}^2}{2m}
        + \tilde{p} + \frac{\hbar^2}{2m\rho_0}|\bn\psi|^2
        \\
        &+ \ii\frac{\hbar\nu}{\rho_0}|\bn\psi|^2  \bigg]
        \tensor{I}\ket{\psi}
        + \vec{\sigma}\cdot\vec{B}\ket{\psi},
    \end{aligned}
\end{equation}
with a modified pressure $\tilde{p}=m(p/\rho_0 - |\vec{u}|^2/2 - \phi_h) + \hbar\vec{s}\cdot\vec{f}/\rho_0$.

Here, $\vec{f}$ is given by Eq.~\eqref{eq:f} with $A_{ij}=\p_is_j$ and $\vec{\mathcal{S}}=2m\nu|\bn\psi|^2\vec{u}/\hbar + m\rho_0\bn\phi_h/\hbar$, and the topology-related scalar $\phi_h$ satisfies
\begin{equation}\label{eq:vor_dot_gradphih}
    \vec{\omega}\cdot\bn\phi_h = -\frac{2\nu}{\rho_0}|\bn\psi|^2h
\end{equation}
with the helicity density $h\equiv \vec{u}\cdot\vec{\omega}$.
Local solutions of $\phi_h$ to Eq.~\eqref{eq:vor_dot_gradphih} can exist~\cite{Greene1993_Reconnection}, but the existence and uniqueness of global solutions to Eq.~\eqref{eq:vor_dot_gradphih} are not guaranteed~\cite{Huang1997_Computation, Hao2019_Tracking}, except that $\vec{f}$ can be obtained by setting $\phi_h=0$ for simple flows with $h=0$.

The NSE-SPE mapping for incompressible flows exhibits a remarkable geometric relation.
It transforms the vorticity, a solenoidal vector field in $\mathbb{R}^3$, into a spin vector in $\mathbb{S}^2$, thereby associating each vortex line in 3D space with a point on the unit (Bloch) sphere~\cite{Chern2017_Inside, Meng2023_Quantum, Meng2024_Lagrangian}.
Note that the spin vector $\vec{s}(\vec{x},t)$, akin to the vorticity field $\vec{\omega}(\vec{x},t)$, is also a smooth vector field in $\mathbb{R}^3$~\cite{Meng2024_Lagrangian}.
From the perspective of vortex dynamics, $\vec{s}$ encodes vortex lines and surfaces~\cite{Chern2016_Schrodinger, Chern2017_Inside, Yang2023_Applications, Meng2024_Lagrangian}. 
Isosurfaces of each component of $\vec{s}$ are a family of vortex surfaces. 
Correspondingly, intersections of isosurfaces of two components of $\vec{s}$ are a family of vortex lines.

The energy expectation of Eq.~\eqref{eq:Schrodinger_4} is
\begin{equation}\label{eq:avgE_incomp}
    \ave{E} = m\rho_0\int_{\mathcal{D}}\bigg(\frac{|\vec{u}|^2}{2} + \frac{p}{\rho_0} - \phi_h \bigg)\dif\vec{x}.
\end{equation}
Note that $\vec{f}$ has an exact expression for 3D compressible flows, whereas in general $\vec{f}$ cannot be determined for incompressible flows.
This implies that incompressible viscous flow corresponds to a peculiar quantum system.
The quantum system in Eq.~\eqref{eq:Schrodinger_4} always has an real-valued energy expectation in Eq.~\eqref{eq:avgE_incomp}, unlike the possible complex values in compressible flows.



For $\nu=0$, Eq.~\eqref{eq:incompressible_NS} reduces to the incompressible Euler equations~\cite{Meng2024_Lagrangian}
\begin{equation}\label{eq:Euler_Eq}
    \begin{dcases}
        \bn\cdot\vec{u} = 0, \\
        \frac{\p\vec{u}}{\p t} + \vec{u}\cdot\bn\vec{u} + \bn\frac{p}{\rho_0} = \vec{0}.
    \end{dcases}
\end{equation}
With $\vec{f}=\vec{0}$ and $\phi_h=0$, the corresponding SPE simplifies to
\begin{equation}\label{eq:Schrodinger_5}
    \ii\hbar\frac{\p}{\p t}\ket{\psi}
    = \bigg(\frac{\hat{\vec{p}}^2}{2m}
    + \tilde{p} + \frac{\hbar^2}{2m\rho_0}|\bn\psi|^2 \bigg)\tensor{I}\ket{\psi}
    + \vec{\sigma}\cdot\vec{B}\ket{\psi}
\end{equation}
with $\tilde{p}=m(p/\rho_0 - |\vec{u}|^2/2)$ and $\vec{B}=\hbar^2\Delta\vec{s}/(4m\rho_0)$.

\subsection{Remarks on the hydrodynamic SPE}
The lengthy derivation for SPEs in Eqs.~\eqref{eq:Schrodinger_2} and \eqref{eq:Schrodinger_4} (see Appendix~\ref{app:derivation_SPE_NSE}) is outlined below.
It begins from the hydrodynamic Schr\"odinger equation~\cite{Chern2016_Schrodinger, Chern2017_Inside, Meng2023_Quantum} for a non-Newtonian fluid flow.
The present SPE adds the imaginary diffusion of the wave function~\cite{Salasnich2024_Quantum} and three terms to be determined, including the real potential that applies a pressure to the fluid, the imaginary potential that balances the imaginary diffusion to ensure the particle-number conservation, and the Stern-Gerlach term $\vec{\sigma}\cdot\vec{B}$ that cancels the artificial body forces (e.g.~the Landau-Lifshitz force~\cite{Meng2023_Quantum}) in the momentum equation.
These undetermined terms are derived by substituting the proposed SPE into the continuity and momentum fluid equations with Eqs.~\eqref{eq:vel} and \eqref{eq:rho}. 
Note that the imaginary diffusion is not the only way to incorporate viscous dissipation within a Hamiltonian framework for fluid dynamics, there are several others such as the trajectory stochasticity~\cite{Eyink2010_Stochastic} and the lattice gas model~\cite{Frisch1986_Lattice, Yepez2001_Quantum}.

To enhance the comprehension of the derivation of the hydrodynamic SPEs, Fig.~\ref{fig:qm-fm_supp} delineates the quantum representation of fluid equations for potential, Schrödinger, Euler, and Navier-Stokes flows.  
The progression of the quantum representation for fluid dynamics, from simple to complex flows, is detailed in Appendix~\ref{app:quantum_repre_steps}.

The SPE and NSE representations have the same number of unknowns.  
For compressible flows, the density and velocity in three directions yield a total of four free variables. 
Notably, the energy equation and the equation of state are computed independently, allowing the pressure to be determined. 
The two-component wave function comprises four real components, i.e., four unknowns.  

For incompressible flows with constant density, the three velocity components satisfy the divergence-free condition, resulting in a total of two unknowns in the NSE representation. 
Meanwhile, the four real components of the wave function (i.e., $a=\mathrm{Re}[\psi_+]$, $b=\mathrm{Im}[\psi_+]$, $c=\mathrm{Re}[\psi_-]$, and $d=\mathrm{Im}[\psi_-]$) satisfy the normalization condition $a^2+b^2+c^2+d^2=\rho_0$ and the divergence-free condition $a\nabla^2b - b\nabla^2a + c\nabla^2d - d\nabla^2c=0$, also yielding two unknowns in the SPE representation.
In summary, the quantum representation aligns with the classical fluid representation in terms of numbers of unknowns.

\subsection{Hamiltonian of the hydrodynamic SPE}\label{sec:Hamiltonian}
The general form $\ii\hbar\p_t\ket{\psi} = \hat{H}\ket{\psi}$ of the SPE for fluid flows has the Hamiltonian
\begin{equation}\label{eq:Hamiltonian}
    \hat{H}
    = \bigg[\frac{\hat{\vec{p}}^2}{2m} + V_r + \ii \bigg(\frac{\hat{\vec{p}}^2}{2m_i} + V_i \bigg)\bigg]\tensor{I}
    + \vec{\sigma}\cdot\vec{B}.
\end{equation}
It depends on the particle mass $m$, virtual mass $m_i$, real potential $V_r$, imaginary potential $V_i$, and the spin-coupled vector $\vec{B}$. 
Given that $V_r$, $V_i$, and $\vec{B}$ in the Hamiltonian can be functions of $\psi$, the SPE is nonlinear in general.
Note that the present SPE does not involve an external electromagnetic field in the standard SPE~\cite{Bransden2003_Physics}.

\begin{table*}[htbp]
    \renewcommand\arraystretch{1.6}
    \caption{Summary of the Hermiticity of the Hamiltonian, virtual mass $m_i$, real potential $V_r$, imaginary potential $V_i$, and spin-coupled vector $\vec{B}$ in Eq.~\eqref{eq:Hamiltonian} for different flows. Note that for the potential flow, $\psi$ degenerates to the one-component wave function.}
    \label{tab:sum_SP}
    \begin{ruledtabular}
        \begin{tabular}{lccccc}
	    Flow                         & Hamiltonian   & $m_i$ & $V_r$ & $V_i$ & $\vec{B}$              \\
            \colrule
    Potential flow~\cite{Madelung1927_Quantentheorie}   & Hermitian &   $\infty$          & Given function    &    $0$   &     --     \\
		Schrödinger flow~\cite{Chern2016_Schrodinger,Meng2023_Quantum} & Hermitian &   $\infty$          & Given function    & $0$       &    $\vec{0}$       \\
		Euler flow~\cite{Meng2024_Lagrangian}           & Hermitian     &    $\infty$        &   $\tilde{p} + \frac{\hbar^2}{2m\rho}|\bn\psi|^2$    &   $0$    &    $\frac{\hbar^2}{4m\rho}\Delta\vec{s} - \hbar\vec{f}$       \\
		Navier-Stokes flow         & non-Hermitian     &    $-\frac{\hbar\rho}{2\mu}$         &   $\tilde{p} + \frac{\hbar^2}{2m\rho}|\bn\psi|^2$    &   $\frac{\hbar\mu}{2}\frac{2|\bn\psi|^2 - \Delta\rho}{\rho^2}$    &    $\frac{\hbar^2}{4m\rho}\Delta\vec{s}-\hbar\vec{f}$
	\end{tabular}
    \end{ruledtabular}
\end{table*}

Table~\ref{tab:sum_SP} presents the expressions of $\hat{H}$ for various flows.
The introduction of the viscous dissipation of Newtonian fluids requires adding the imaginary diffusion~\cite{Salasnich2024_Quantum} of $\psi$ with a virtual mass $m_i$, which makes the quantum system non-Hermitian.
To ensure the probability conservation, we invoke the imaginary potential $V_i$.
Hence, despite the non-Hermiticity of the Hamiltonian in the viscous flow, the quantum state evolves unitarily, enabling its potential implementation on a gate-based quantum computer.
The anti-Hermitian term $\ii\hat{H}_D$~\cite{Sone2019_Anomalous, Ashida2021_Non}, with $\hat{H}_D=\hat{H}_D^\dagger$ and $\hat{H}_D=\hat{\vec{p}}^2/(2m_i)+V_i$, arises from the viscous dissipation in fluids.
The virtual mass, proportional to $1/\nu$, has $m_i\rightarrow\infty$ in the inviscid flow without the non-Hermitian effect.

As summarized in Table~\ref{tab:sum_SP}, simplifying the Hamiltonian corresponds to a quantum representation of simpler flows, which can lead to various applications.
For the Euler (inviscid) flow, the imaginary diffusion and potential are removed from the SPE.  
The simplified SPE in Eq.~\eqref{eq:Schrodinger_5} 
corresponds to the Euler equation in Eq.~\eqref{eq:Euler_Eq}. 
In particular, the SPE for incompressible Euler flow offers a unique Lagrangian perspective to study vortex-surface evolution \cite{Meng2024_Lagrangian}.
%
For the Schrödinger flow, which has been used in quantum simulation \cite{Meng2024_Simulating} and computer graphics \cite{Chern2016_Schrodinger} for fluid dynamics, the spin-coupled term $\vec{\sigma}\cdot\vec{B}\ket{\psi}$ is removed from Eq.~\eqref{eq:Schrodinger_5}. 
For the potential flow, the SPE degenerates to the standard time-dependent Schrödinger equation, and the NSE-SPE mapping degenerates to the classical Madelung transform~\cite{Madelung1927_Quantentheorie}. 

\section{Quantum representation for flow evolution}\label{sec:SPE_evolution}
We develop a numerical algorithm to solve Eq.~\eqref{eq:Schrodinger_2} for compressible flows and Eq.~\eqref{eq:Schrodinger_4} for incompressible, constant-density flows with $\rho_0=1$.
The solution domain $\mathcal{D}=[0,2\pi]^d$ with the space dimension $d$ and periodic boundary condition is discretized on $N^d$ uniform grid points.
We employ the fraction-step Fourier method based on the second-order Trotter-Suzuki decomposition~\cite{Suzuki1990_Fractal, Suzuki1991_General} for time integration.
For compressible flows, we directly compute the modified pressure by integrating the static pressure from the equation of state.
For incompressible flows, we project a temporary wave function onto a divergence-free subspace, instead of solving the pressure-Poisson equation to calculate the pressure.

The numerical method is presented in Algorithm~\ref{al:SPE_NS} and the code for simulating the SPE is available~\cite{code}, where $\vec{k}=(k_1,\cdots,k_d)$ is the $d$-dimensional wavenumber vector with $k_i=-N/2,-N/2+1,\cdots,N/2-1,~i=1,2,\cdots,d$, $k=|\vec{k}|$ is the wavenumber magnitude, and $\mathcal{F}$ and $\mathcal{F}^{-1}$ represent the Fourier transform and its inverse, respectively.
For brevity, the normalization constant of the wave function is omitted in the numerical simulation, such that $\int_\mathcal{D}\langle\psi|\psi\rangle\dif\vec{x}\ne 1$.
Note that we only address the periodic boundary condition in this paper. 
The treatment of general boundary conditions (e.g., solid walls and free surfaces) for the wave function can be intricate~\cite{Yang2021_Clebsch}.

\begin{algorithm}
    \caption{Time iteration for solving the SPE for fluid flows.}
    \label{al:SPE_NS}
    \KwIn{$\ket{\psi(\vec{x},t)}$, $\mu$ (or $\nu$), $\dt$, $N$}
    \KwOut{$\ket{\psi(\vec{x},t+\dt)}$}
    Calculate the spin vector $\vec{s}(\vec{x},t)$\;
    \If{compressible}{
    Calculate the fluid density $\rho(\vec{x},t)$ and velocity $\vec{u}(\vec{x},t)$\;
    Calculate the coefficient matrix $\tensor{A}(\vec{x},t)$, the source term $\vec{\mathcal{S}}(\vec{x},t)$, the force $\vec{f}(\vec{x},t)$, and the modified pressure $\tilde{p}(\vec{x},t)$\;
    Calculate the real and imaginary potentials $V_r(\vec{x},t)=|\bn\psi|^2/(2\rho)$ and $V_i(\vec{x},t)=\mu(2|\bn\psi|^2 - \Delta\rho)/(2\rho^2)$, respectively\;
    Calculate the spin-coupled potentials $P(\vec{x},t)=\Delta s_3/(4\rho) - f_3$ and $Q(\vec{x},t)=(\Delta s_1 - \ii\Delta s_2)/(4\rho) - (f_1 - \ii f_2)$\;
    $\psi_\pm^*(\vec{x}) \leftarrow \mathcal{F}^{-1}\{\ee^{-(\mu/\rho + \ii/2)k^2\dt}\mathcal{F}\{\psi_\pm\} \}$\;
    $\ket{\psi(\vec{x},t+\dt)} \leftarrow \ee^{-\ii(\tilde{p} + V_r \pm P + \ii V_i)\dt}\psi_\pm^* + \big(\ee^{-\ii(\mathrm{Re}[Q] \pm \ii\mathrm{Im}[Q])\dt} - 1\big)\psi_\mp^*$.
    }
    \If{incompressible}{
    Calculate the fluid velocity $\vec{u}(\vec{x},t)$\;
    Calculate the coefficient matrix $\tensor{A}(\vec{x},t)$, the topology-related scalar $\phi_h(\vec{x},t)$, the source term $\vec{\mathcal{S}}(\vec{x},t)$, and the force $\vec{f}(\vec{x},t)$\;
    Calculate the real and imaginary potentials $V_r(\vec{x},t)=|\bn\psi|^2/2$ and $V_i(\vec{x},t)=\nu|\bn\psi|^2$, respectively\;
    Calculate the spin-coupled potentials $P(\vec{x},t)=\Delta s_3/4 - f_3$ and $Q(\vec{x},t)=(\Delta s_1 - \ii\Delta s_2)/4 - (f_1 - \ii f_2)$\;
    $\psi_\pm^*(\vec{x}) \leftarrow \mathcal{F}^{-1}\{\ee^{-(\nu + \ii/2) k^2\dt}\mathcal{F}\{\psi_\pm\}\}$\;
    $\psi_\pm^{**}(\vec{x}) \leftarrow \ee^{-\ii(V_r \pm P + \ii V_i)\dt}\psi_\pm^* + \big(\ee^{-\ii(\mathrm{Re}[Q] \pm \ii\mathrm{Im}[Q])\dt} - 1 \big)\psi_\mp^*$\;
    $\psi_\pm^{***}(\vec{x}) \leftarrow \psi_\pm^{**} / |\psi_\pm^{**}|$\;
    Obtain the phase shift $q(\vec{x},t)$ by solving the Poisson equation $\Delta q=-\mathrm{Im}\langle\psi^{***}|\hat{\vec{p}}^2|\psi^{***}\rangle/\hbar^2$\;
    $\ket{\psi(\vec{x},t+\dt)} \leftarrow \ee^{-\ii q}\ket{\psi^{***}(\vec{x})}$.
    }
\end{algorithm}

Next, we illustrate the quantum/wave-like behavior in the flow evolution using the SPE-based numerical simulation for a 1D compressible flow and a 2D incompressible Taylor-Green (TG) flow~\cite{Taylor1937_Mechanism}.
Without loss of generality, we set $\rho_0=1$, $\hbar=1$, and $m=1$.

\subsection{1D Burgers equation}
We consider a 1D compressible flow governed by the Burgers equation at $x\in[0,2\pi]$ with the initial condition $u(x,0)=\sin x$, boundary condition $u(0,t)=u(2\pi,t)=0$, and $\nu=0.01$.
The initial wave function is $\psi_+(x,0)=\cos x\ee^{-\ii\cos x}$ and $\psi_-(x,0)=\sin x\ee^{-\ii \cos x}$.
The SPE formulation and the corresponding numerical algorithm of 1D Burgers equation are detailed in Appendix~\ref{app:SPE_Burgers}.

\begin{figure}
    \centering
    \includegraphics[width=8.6cm]{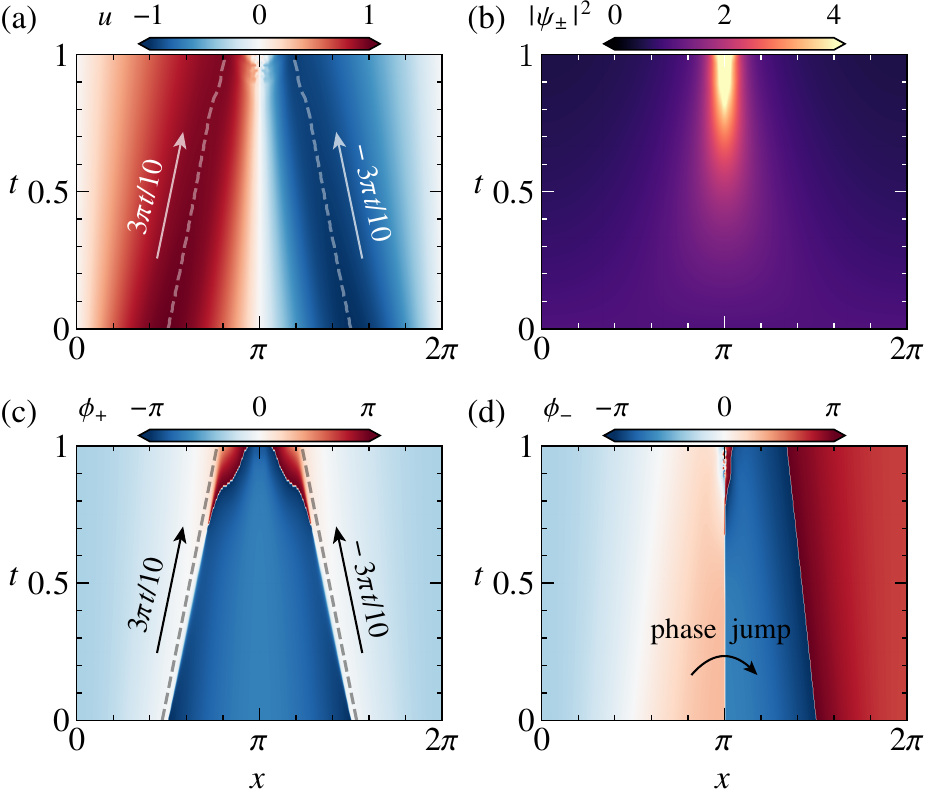}
    \caption{Numerical solutions to the SPE for the Burgers equation:
    (a) velocity reconstructed from Eq.~\eqref{eq:vel},
    (b) particle's probability density, and
    (c, d) phases of spin-up and -down components.}
    \label{fig:Burgers_rho_U}
\end{figure}

\begin{figure}
    \centering
    \includegraphics{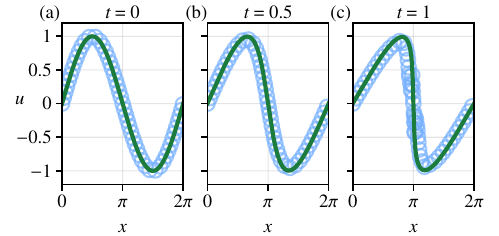}
    \caption{Comparison of the reconstructed velocity (circles) obtained by numerically solving the SPE equivalent to the Burgers equation with the exact solution (lines), at $t=0$, 0.5, and 1.}
    \label{fig:Burger_u_compare}
\end{figure}

\begin{figure}
    \centering
    \includegraphics{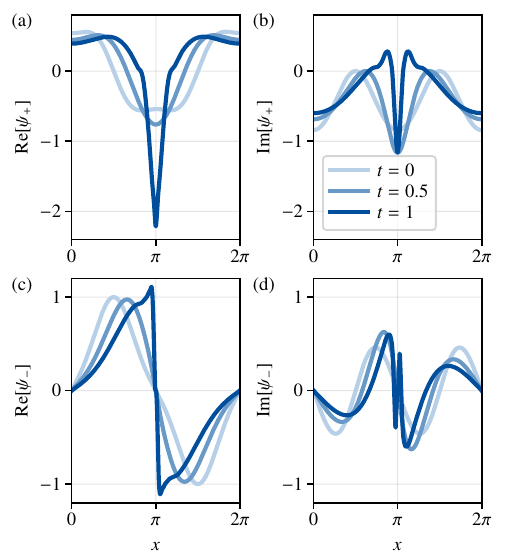}
    \caption{Evolution of the real and imaginary parts (a) $\mathrm{Re}[\psi_+]$, (b) $\mathrm{Im}[\psi_+]$, (c) $\mathrm{Re}[\psi_-]$, and (d) $\mathrm{Im}[\psi_-]$ of $\psi_+$ and $\psi_-$, obtained by numerically solving the SPE equivalent to the Burgers equation.}
    \label{fig:Burger_psi}
\end{figure}

This flow is a simplified model for the formation of a shock wave.
In Figs.~\ref{fig:Burgers_rho_U}(a) and \ref{fig:Burger_u_compare}, the reconstructed velocity from the numerical solution to Eq.~\eqref{eq:vel} agrees with the exact solution
\begin{equation}
    u(x,t) = -2\nu\frac{\sum_{k=-\infty}^\infty\ii k\ee^{-\nu k^2t + \ii kx}\int_0^{2\pi}\ee^{\cos x'/(2\nu)}\ee^{-\ii kx'}\dif x'}{\sum_{k=-\infty}^\infty \ee^{-\nu k^2t + \ii kx}\int_0^{2\pi}\ee^{\cos x'/(2\nu)}\ee^{-\ii kx'}\dif x'}.
\end{equation}
The fluids on both sides converge to the center at $x=\pi$, forming a shock.
Note that the numerical solution is only shown at $t\le 1$, due to strong numerical oscillations in solving the SPE at later times.
The numerical oscillations are particularly evident in Figs.~\ref{fig:Burger_u_compare}(c) and \ref{fig:Burger_psi}(d), as indicated by the pronounced overshooting of the solution near $x = \pi$.
This could be attributed to the dispersion error in discretizing the wave function's high-frequency components.
In Fig.~\ref{fig:Burgers_rho_U}(b), the probability of particle occurrence accumulates near the shock.
Thus, the velocity discontinuity corresponds to the maximum probability amplitude in the wave-function representation.
Besides, the temporal evolution of the real and imaginary parts of $\psi$ in Fig.~\ref{fig:Burger_psi} shows notable gradient magnitudes of the wave function at $x=\pi$, suggesting the shock formation. 

The quantum representation can shed light on the flow evolution.
Figures~\ref{fig:Burgers_rho_U}(c) and (d) plot the phases $\phi_\pm\equiv \arg(\psi)$ of spin-up and -down components, respectively.
The contour of $\phi_+$ exhibits two lines separating large and near-zero values, indicating two converging waves with the speed of $\pm 3\pi/10$. They coincide with the maximum-speed locations (dashed lines in Fig.~\ref{fig:Burgers_rho_U}(a) and (c)), and defining the shock emergence at $t=5/3$.
Furthermore, the phase jump $\phi_-(\pi^-,t)-\phi_-(\pi^+,t)=\pi$ at the initial time in Fig.~\ref{fig:Burgers_rho_U}(d) signals the shock location $x=\pi$ at later times.

\subsection{2D incompressible Taylor-Green viscous flow}
We consider the 2D incompressible TG viscous flow in a periodic box $\vec{x}\in[0,2\pi]^2$ with $\vec{u}(\vec{x},0)=\sin x\cos y\,\vec{e}_x - \cos x\sin y\,\vec{e}_y$ and $\nu=1$, where $\{\vec{e}_x, \vec{e}_y\}$ are the Cartesian unit vectors.
The initial wave function is $\psi_+(\vec{x},0)=\cos[H(x)]\ee^{\ii\cos y(2-\cos x)}$ and $\psi_-(\vec{x},0)=\sin[H(x)]\ee^{-\ii\cos y(2+\cos x)}$, with $H(x)=x/2$ for $0\le x\le \pi$ and $H(x)=\pi - x/2$ for $\pi<x\le 2\pi$.
The exact solutions for the velocity and vorticity are $\vec{u}(\vec{x},t)=\ee^{-2\nu t}(\sin x\cos y\vec{e}_x - \cos x\sin y\vec{e}_y)$ and $\omega_z(\vec{x},t)=2\ee^{-2\nu t}\sin x\sin y$, respectively.

\begin{figure}
    \centering
    \includegraphics[width=8.6cm]{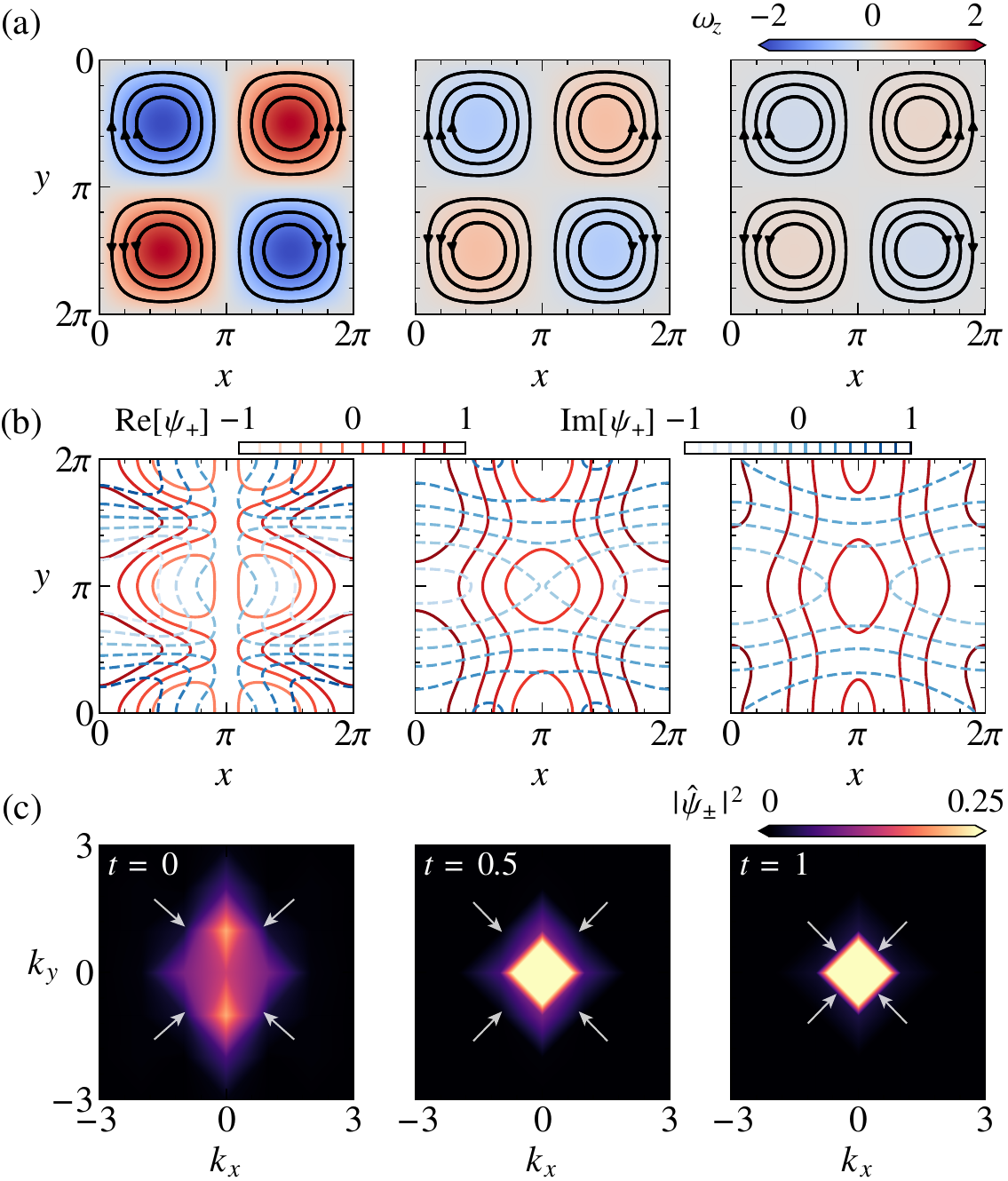}
    \caption{Numerical results obtained from the simulation of Eq.~\eqref{eq:Schrodinger_4} for the 2D TG flow at $t=0$ (left), $0.5$ (middle), and $1$ (right).
    (a) Streamlines and vorticity contour.
    (b) Contours of the real (solid red curves) and imaginary (dashed blue curves) parts of the spin-up component.
    (c) Particle probability distribution in the momentum space.}
    \label{fig:2DTG}
\end{figure}

\begin{figure}
    \centering
    \includegraphics{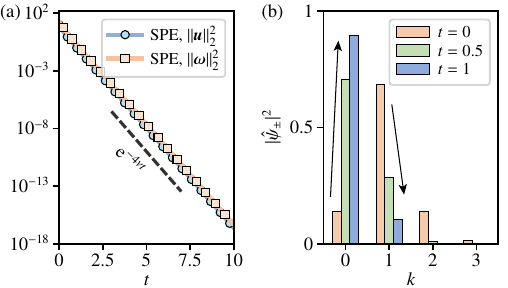}
    \caption{(a) Total kinetic energy $\norm{\vec{u}}_2^2$ and enstrophy $\norm{\vec{\omega}}_2^2$ of the 2D TG flow simulated by Eq.~\eqref{eq:Schrodinger_4}. Both decay exponentially as predicted by the exact solution.
    (b) Momentum representation of the wave-function probability distribution at $t=0$, $0.5$, and $1$. The particle evolves towards the zero momentum state with time.}
    \label{fig:2DTG_statistics}
\end{figure}

This flow illustrates the quantum representation for the viscous effect.
The numerical result exhibits an excellent agreement with the exact solution in Figs.~\ref{fig:2DTG}(a) and \ref{fig:2DTG_statistics}(a) -- the total kinetic energy $\norm{\vec{u}}_2^2$ and enstrophy $\norm{\vec{\omega}}_2^2$ decay exponentially.
The isolines of the real (solid red lines) and imaginary (dashed blue lines) parts of $\psi_+$ at $t=0$, $0.5$, and $1$ are plotted Fig.~\ref{fig:2DTG}(b).
They are interlaced at the initial time, and then elongate along the $x$- and $y$-directions in the viscous decaying process.
In the pilot wave theory~\cite{Bohm1952_A, Takabayasi1952_On, Takabayasi1953_Remarks, Wyatt2005_Quantum, Bohm2012_Quantum}, the smoothing of $\psi$ reduces the velocity of fluid particles, consistent with the observed attenuation of vortex rotation in Fig.~\ref{fig:2DTG}(a).
The particle probability distribution in the momentum space converges to the zero state in Fig.~\ref{fig:2DTG}(c), implying that the fluid velocity decays to zero under viscous dissipation.
In Fig.~\ref{fig:2DTG_statistics}(b), the peak of the wave-function probability distribution migrates towards the zero-momentum state with time.

Note that the nonlinear SPE can display significant numerical oscillations (see Figs.~\ref{fig:Burgers_rho_U}, \ref{fig:Burger_u_compare}(c) and \ref{fig:Burger_psi}(d)) when solved using the fraction-step Fourier method without effective damping techniques.
Therefore, it is essential to devise a numerical method for mitigating these oscillations in solving the quantum representation of turbulent flows.

\section{Discussion}\label{sec:discussion}
We develop a quantum representation for Navier-Stokes fluid dynamics, establishing a mapping between the hydrodynamic SPE in Eqs.~\eqref{eq:Schrodinger_2} and \eqref{eq:Schrodinger_4} for the compressible and incompressible NSEs, respectively.
Beyond the classical Madelung transform~\cite{Madelung1927_Quantentheorie}, the present SPE describes a quantum spin system with a two-component spinor wave function, corresponding to a flow with finite vorticity.
The wave function undergoes imaginary diffusion~\cite{Salasnich2024_Quantum} to account for the viscous dissipation of Newtonian fluids, rendering the quantum system non-Hermitian.
To preserve probability conservation, an imaginary potential is introduced.

On the other hand, a globally smooth wave function may not exist for a velocity field with vorticity nulls or unclosed vortex lines~\cite{Meng2024_Lagrangian}.
Therefore, the SPE is limited to the NSE with a subset of initial conditions.
A useful approximation of the wave function can be obtained for a given velocity field using the numerical optimization~\citep{Chern2017_Inside, Su2024_Quantum}.
Note that an additional constraint equation in Eq.~\eqref{eq:vor_dot_gradphih} must be solved for 3D incompressible viscous flows with $h\ne 0$. While local solutions to Eq.~\eqref{eq:vor_dot_gradphih} can exist, the existence and uniqueness of global solutions are not guaranteed.

Moreover, the nonlinear SPE exhibits significant numerical oscillations when solved using the fraction-step Fourier method without effective damping techniques for flows with strong nonlinearity. 
Consequently, it is crucial to develop a numerical method to mitigate these oscillations when solving the quantum representation of turbulent flows.

We demonstrate the quantum/wave-like behavior of the flow evolution for a 1D compressible flow and a 2D incompressible TG flow using the SPE-based numerical simulation.
In the compressible flow, the wave-function representation exhibits a maximum particle probability at the velocity discontinuity near the shock, and the initial phase jump signals the location of shock formation at later times.
In the incompressible flow, the real and imaginary parts of the wave function are initially interlaced, and then elongate along the $x$- and $y$-directions as the flow undergoes viscous dissipation.
The particle probability distribution in the momentum space approaches the zero state, reflecting the quantum representation of viscous effects.

The present study can open new research avenues in the interdisciplinary study of fluid dynamics and quantum mechanics.
First, we demonstrate the potential of quantum simulation~\cite{Feynman1982_Simulating} for classical fluid dynamics, including viscous and vortical flows.
The SPE, a Hamiltonian system, is more tractable than the NSE for quantum computing~\cite{Meng2023_Quantum}, and its simplified form has been used in the unsteady flow simulation on a NISQ device~\cite{Meng2024_Simulating}.
The synergy of computational fluid dynamics and quantum computing heralds the next-generation simulation method~\cite{Givi2020_Quantum, Succi2023_Quantum, Bharadwaj2023_Hybrid}.

However, there are two primary obstacles that need to be addressed -- the nonlinearity and the non-Hermitian Hamiltonian in the SPE.
The mean-field nonlinear quantum algorithm~\cite{Lloyd2020_Quantum, Großardt2024_Nonlinear, Brüstle2024_Quantum} partially addresses the former by employing symmetric interactions among $n\gg 1$ identical quantum-state copies to approximate the nonlinear evolution of an individual copy. 
The mean-field linearization provides an alternative approach, converting the nonlinear potential acting on a single particle into a linear interaction term within a quantum many-body system.
For the latter, since the quantum state evolution maintains unitary and the non-Hermitian part is anti-Hermitian~\cite{Ashida2021_Non}, it seems feasible to implement this Hamiltonian on a gate-based quantum computer by introducing auxiliary variables~\cite{Jin2023_Quantum}.
Nevertheless, devising an effective quantum algorithm for evolving this quantum system remains a valuable open problem.

Second, the hydrodynamic SPE enables studying classical turbulence as a quantum system, thereby extending previous research on quantum turbulence~\cite{Muller2021_Intermittency, Shen2024_Weaving}.
This approach offers a unique perspective, leveraging the inherent Lagrangian nature of the quantum representation to facilitate the investigation of vortex-surface evolution in incompressible flows~\cite{Yang2023_Applications, Meng2024_Lagrangian}.
Furthermore, the NSE-SPE correspondence may unveil novel quantum effects on classical mechanical systems within the framework of non-Hermitian physics~\cite{Ashida2021_Non, Bergholtz2021_Exceptional, Ding2022_Non}.

\begin{acknowledgments}
Numerical simulations were carried out on the Tianhe-2A supercomputer in Guangzhou, China.
This work has been supported by the National Natural Science Foundation of China (Grant Nos.~11925201 and 11988102), the National Key R\&D Program of China (Grant Nos.~2023YFB4502600 and 2020YFE0204200), and the Xplore Prize.
\end{acknowledgments}

\appendix

\section{Schr\"odinger-Pauli formulation of the NSE}\label{app:derivation_SPE_NSE}

\subsection{Derivation of the SPE for 3D compressible flows}
We derive the SPE in Eq.~\eqref{eq:Schrodinger_2} which is equivalent to the NSE in Eq.~\eqref{eq:compressible_NS} for general 3D flows.
To incorporate the finite vorticity $\vec{\omega}\equiv \bn\times\vec{u}$ into the hydrodynamic representation of the SPE,
we employ a two-component wave function as a quaternion
\begin{equation}
    \vec{\psi}(\vec{x},t) = a(\vec{x},t) + b(\vec{x},t)\vec{i} + c(\vec{x},t)\vec{j} + d(\vec{x},t)\vec{k}
\end{equation}
with the basis vectors $\{\vec{i},\vec{j},\vec{k}\}$ of the imaginary part of the quaternion and real-valued functions $a$, $b$, $c$, and $d$.
Note that this quaternion, equivalent to the two-component spinor, facilitates deriving the hydrodynamic SPE below.

The probability current is defined as~\cite{Meng2023_Quantum}
\begin{equation}
    \vec{J} \equiv \frac{\hbar}{2m}(\bn\widebar{\vec{\psi}}\vec{i\psi} - \widebar{\vec{\psi}}\vec{i}\bn\vec{\psi})
\end{equation}
where $\hbar$ is the reduced Planck constant, $m$ the particle mass, and $\widebar{\vec{\psi}}$ the quaternion conjugate of $\vec{\psi}$.
The flow velocity is then defined as
\begin{equation}\label{eq:SM_vel}
    \vec{u} \equiv \frac{\vec{J}}{\rho}
    = \frac{\hbar}{2m}\frac{\bn\widebar{\vec{\psi}}\vec{i\psi} - \widebar{\vec{\psi}}\vec{i}\bn\vec{\psi}}{\widebar{\vec{\psi}}\vec{\psi}},
\end{equation}
where $\rho\equiv\widebar{\vec{\psi}}\vec{\psi}$ is the fluid density.

We propose a nonlinear SPE that is equivalent to the NSE as
\begin{equation}\label{eq:SM_Schrodinger_0}
    \vec{i}\hbar\frac{\p\vec{\psi}}{\p t} = -\frac{\hbar^2}{2m}(1-\vec{i}\nu')\Delta\vec{\psi} + \vec{V}\vec{\psi}
    - \frac{\hbar^2}{4m\rho}\vec{i\psi}\Delta\tilde{\vec{s}} + \hbar\vec{i\psi f},
\end{equation}
where $\Delta$ is the Laplacian operator, $\nu'>0$ serves as an effective viscosity, $\vec{V}=V_1+V_2\vec{i}$ is a quaternion with two real-valued functions $V_1$ and $V_2$ to be determined, $\tilde{\vec{s}}\equiv\widebar{\vec{\psi}}\vec{i\psi}=(a^2+b^2-c^2-d^2)\vec{i}+2(bc-ad)\vec{j}+2(ac+bd)\vec{k}$ is a pure quaternion for the spin vector in $\mathbb{R}^3$, and $\vec{f}=f_1\vec{i}+f_2\vec{j}+f_3\vec{k}$ is a pure quaternion to be determined.

The choice of the present SPE form in Eq.~\eqref{eq:SM_Schrodinger_0} is explained below.
The first term on the right-hand side (RHS) involves the imaginary diffusion~\cite{Salasnich2024_Quantum}, which accounts for the proper viscous dissipation in Newtonian fluids.
The $\vec{i}$-component $V_2$ of the potential $\vec{V}$ is employed to preserve the particle number.
The last two terms on the RHS of Eq.~\eqref{eq:SM_Schrodinger_0} cancel the artificial body force, e.g., the Landau-Lifshitz force~\cite{Meng2023_Quantum}, in the momentum equation.

Combining Eqs.~\eqref{eq:SM_vel} and \eqref{eq:SM_Schrodinger_0}, we have
\begin{equation}\label{eq:continuty}
\begin{aligned}
    &\frac{\p\rho}{\p t} + \bn\cdot(\rho\vec{u})
    \\
    =\ & \bigg(\frac{\p\widebar{\vec{\psi}}}{\p t} + \frac{\hbar}{2m}\Delta\widebar{\vec{\psi}}\vec{i} \bigg)\vec{\psi}
    + \widebar{\vec{\psi}}\bigg(\frac{\p\vec{\psi}}{\p t} - \frac{\hbar}{2m}\vec{i}\Delta\vec{\psi} \bigg)
    \\
    =\ & \bigg(\frac{\hbar\nu'}{2m}\Delta\widebar{\vec{\psi}} + \frac{1}{\hbar}\widebar{\vec{\psi}}\widebar{\vec{V}}\vec{i} \bigg)\vec{\psi}
    + \widebar{\vec{\psi}}\bigg(\frac{\hbar\nu'}{2m}\Delta\vec{\psi} - \frac{1}{\hbar}\vec{iV\psi} \bigg)
    \\
    =\ & \frac{\hbar\nu'}{2m}(\Delta\widebar{\vec{\psi}}\vec{\psi} + \widebar{\vec{\psi}}\Delta\vec{\psi})
    + \frac{1}{\hbar}\widebar{\vec{\psi}}(\widebar{\vec{V}}\vec{i} - \vec{iV})\vec{\psi}
    \\
    =\ & \frac{\hbar\nu'}{2m}\left(\Delta\rho - 2|\bn\vec{\psi}|^2 \right) + \frac{2}{\hbar}\rho V_2.
\end{aligned}
\end{equation}
To satisfy the continuity equation $\p\rho/\p t + \bn\cdot(\rho\vec{u})=0$, we impose
\begin{equation}\label{eq:V_2}
    V_2
    = \frac{\hbar^2\nu'}{4m}\frac{2|\bn\vec{\psi}|^2 - \Delta\rho}{\rho}.
\end{equation}

Then, we derive the momentum equation corresponding to Eq.~\eqref{eq:SM_Schrodinger_0}.
Taking the material derivative $\DD/\DD t\equiv \p / \p t + \vec{u}\cdot\bn$ of $\vec{J}$ and using $\vec{J}=(\hbar/m)\mathrm{Re}[{\bn\widebar{\vec{\psi}}\vec{i\psi}}]$ yield
\begin{equation}\label{eq:DJDt_1}
    \frac{\DD\vec{J}}{\DD t}
    = \frac{\hbar}{m}\mathrm{Re}\bigg[\frac{\DD\bn\widebar{\vec{\psi}}}{\DD t}\vec{i\psi} + \bn\widebar{\vec{\psi}}\vec{i}\frac{\DD\vec{\psi}}{\DD t} \bigg].
\end{equation}
Substituting the vector identity
\begin{equation}
    \frac{\DD\bn\widebar{\vec{\psi}}}{\DD t}
    = \bn\frac{\DD\widebar{\vec{\psi}}}{\DD t} - \bn\vec{u}\cdot\bn\widebar{\vec{\psi}}
\end{equation}
into Eq.~\eqref{eq:DJDt_1}, we obtain
\begin{equation}\label{eq:DJDt_2}
    \frac{\DD\vec{J}}{\DD t}
    = \frac{\hbar}{m}\mathrm{Re}\bigg[\bn\frac{\DD\widebar{\vec{\psi}}}{\DD t}\vec{i\psi} + \bn\widebar{\vec{\psi}}\vec{i}\frac{\DD\vec{\psi}}{\DD t}\bigg] - \rho\bn\frac{|\vec{u}|^2}{2}.
\end{equation}
After some algebra, the convection of wave function reads
\begin{equation}\label{eq:convect_psi1}
    \begin{aligned}
        \vec{u}\cdot\bn\vec{\psi}
        =\ & -\frac{\hbar}{2m}\vec{i}\Delta\vec{\psi} - \frac12(\bn\cdot\vec{u})\vec{\psi} 
        \\
        &+ \frac{\hbar}{4m\rho}\left(2|\bn\vec{\psi}|^2\vec{i\psi} - \frac{\vec{\psi}}{\rho}\bn\rho\cdot\bn\tilde{\vec{s}} + \vec{\psi}\Delta\tilde{\vec{s}}\right).
    \end{aligned}
\end{equation}
Combining Eqs.~\eqref{eq:SM_Schrodinger_0} and \eqref{eq:convect_psi1}, we have
\begin{equation}\label{eq:DpsiDt_1}
    \begin{aligned}
        \frac{\DD\vec{\psi}}{\DD t}
        =\ & \frac{\hbar\nu'}{2m}\Delta\vec{\psi} - \frac{1}{\hbar}\vec{iV\psi} - \frac{\hbar}{4m\rho}\vec{\psi}\Delta\tilde{\vec{s}} + \vec{\psi f} - \frac{1}{2}(\bn\cdot\vec{u})\vec{\psi}
        \\
        &+ \frac{\hbar}{4m\rho}\left(2|\bn\vec{\psi}|^2\vec{i\psi} - \frac{\vec{\psi}}{\rho}\bn\rho\cdot\bn\tilde{\vec{s}} + \vec{\psi}\Delta\tilde{\vec{s}}\right).
    \end{aligned}
\end{equation}

Substituting Eq.~\eqref{eq:DpsiDt_1} with its complex conjugate into Eq.~\eqref{eq:DJDt_2} yields
\begin{equation}\label{eq:DJDt_3}
    \begin{aligned}
        \frac{\DD\vec{J}}{\DD t}
        =\ & -\rho(\bn\cdot\vec{u})\vec{u} - \frac{\hbar^2}{4m^2}\bn\left(\frac{1}{\rho}|\bn\rho|^2 \right) 
        + \frac{2V_2\rho}{\hbar}\vec{u}
        + \frac{\hbar}{m}\bn\vec{f}\cdot\tilde{\vec{s}}
        \\
        &+ \frac{\hbar^2}{4m^2\rho^2}\bn\rho\cdot\left[\bn(\rho\bn\rho) - \bn\bn\tilde{\vec{s}}\cdot\tilde{\vec{s}} \right]
        - \frac{\rho}{m}\bn V_1
        \\
        &+ \frac{{\hbar}^2}{2m^2}\rho\bn\frac{|\bn\vec{\psi}|^2}{\rho}
        + \frac{\hbar}{2m}(\rho\bn\cdot\vec{u} + \vec{u}\cdot\bn\rho)\bn\nu'
        \\
        &- \rho\bn\frac{|\vec{u}|^2}{2}
        + \frac{\hbar^2\nu'}{2m^2} \mathrm{Re}\left[\bn\Delta\widebar{\vec{\psi}}\vec{i\psi} + \bn\widebar{\vec{\psi}}\vec{i}\Delta\vec{\psi}  \right].
    \end{aligned}
\end{equation}
Taking the Laplacian of $\vec{J}$, we have
\begin{equation}\label{eq:Laplacian_J}
    \begin{aligned}
        \Delta\vec{J}
        &= \frac{\hbar}{2m}(\bn\Delta\widebar{\vec{\psi}}\vec{i\psi} + \bn\widebar{\vec{\psi}}\vec{i}\Delta\vec{\psi} + 2\bn\bn\widebar{\vec{\psi}}\cdot\vec{i}\bn\vec{\psi} - 2\bn\widebar{\vec{\psi}}\cdot\vec{i}\bn\bn\vec{\psi} 
        \\
        &\hspace{3em}- \Delta\widebar{\vec{\psi}}\vec{i}\bn\vec{\psi} - \widebar{\vec{\psi}}\vec{i}\bn\Delta\vec{\psi}).
    \end{aligned}
\end{equation}
Taking the real part of Eq.~\eqref{eq:Laplacian_J} and using the identities
\begin{equation}
\begin{cases}
\begin{aligned}
    \mathrm{Re}\left[\bn\Delta\widebar{\vec{\psi}}\vec{i\psi} \right]
    =\ & -b\bn\Delta a + a\bn\Delta b - d\bn\Delta c + c\bn\Delta d,
    \\
    \mathrm{Re}\left[\bn\widebar{\vec{\psi}}\vec{i}\Delta\vec{\psi} \right]
    =\ & -\Delta b\bn a + \Delta a\bn b - \Delta d\bn c + \Delta c\bn d,
    \\
    \mathrm{Re}\left[\Delta\widebar{\vec{\psi}}\vec{i}\bn\vec{\psi} \right]
    =\ & -\Delta a\bn b + \Delta b\bn a - \Delta c\bn d + \Delta d\bn c,
    \\
    \mathrm{Re}\left[\widebar{\vec{\psi}}\vec{i}\bn\Delta\vec{\psi} \right]
    =\ & b\bn\Delta a - a\bn\Delta b + d\bn\Delta c - c\bn\Delta d,
    \\    
    \mathrm{Re}\left[\bn\bn\widebar{\vec{\psi}}\cdot\vec{i}\bn\vec{\psi} \right]
    =\ & -\bn b\cdot\bn\bn a + \bn a\cdot\bn\bn b 
    \\
    &- \bn d\cdot\bn\bn c 
    + \bn c\cdot\bn\bn d,
    \\
    \mathrm{Re}\left[\bn\widebar{\vec{\psi}}\cdot\vec{i}\bn\bn\vec{\psi} \right]
    =\ & -\bn a\cdot\bn\bn b + \bn b\cdot\bn\bn a 
    \\
    &- \bn c\cdot\bn\bn d 
    + \bn d\cdot\bn\bn c,
\end{aligned}
\end{cases}
\end{equation}
we obtain
\begin{equation}\label{eq:Laplacian_J_1}
    \Delta\vec{J}
    = \mathrm{Re}[\Delta\vec{J}]
    = \frac{\hbar}{m}\mathrm{Re}\left[\bn\Delta\widebar{\vec{\psi}}\vec{i\psi} + \bn\widebar{\vec{\psi}}\vec{i}\Delta\vec{\psi}\right].
\end{equation}
Substituting the identity
\begin{equation}\label{eq:Laplacian_J_2}
    \Delta\vec{J}
    = \Delta\rho\vec{u} + 2\bn\rho\cdot\bn\vec{u} + \rho\Delta\vec{u}
\end{equation}
and Eqs.~\eqref{eq:V_2} and \eqref{eq:Laplacian_J_1} into Eq.~\eqref{eq:DJDt_3} yields
\begin{equation}\label{eq:Laplacian_J_3}
    \begin{aligned}
        \frac{\DD\vec{J}}{\DD t}
        =\ & - \frac{\hbar^2}{4m^2}\bn\left(\frac{1}{\rho}|\bn\rho|^2 \right)
        + \rho\bn\left(\frac{\hbar^2}{2m^2\rho}|\bn\vec{\psi}|^2 - \frac{|\vec{u}|^2}{2} - \frac{V_1}{m} \right)
        \\        
        &+ \frac{\hbar\nu'}{2m}\rho\Delta\vec{u}
        + \frac{\hbar}{m}\bn\vec{f}\cdot\tilde{\vec{s}}
        + \left(\frac{\hbar\nu'}{m}|\bn\vec{\psi}|^2 - \rho\bn\cdot\vec{u} \right)\vec{u}
        \\
        &+ \frac{\hbar^2}{4m^2\rho^2}\bn\rho\cdot\left[\bn(\rho\bn\rho) - \bn\bn\tilde{\vec{s}}\cdot\tilde{\vec{s}} + \frac{4m\nu'}{\hbar}\rho^2\bn\vec{u} \right]
        \\
        &+ \frac{\hbar}{2m}(\rho\bn\cdot\vec{u} + \vec{u}\cdot\bn\rho)\bn\nu'.
    \end{aligned}
\end{equation}

Substituting Eq.~\eqref{eq:Laplacian_J_3} into the relation $\DD\vec{u}/\DD t = (1/\rho)\DD\vec{J}/\DD t + (\bn\cdot\vec{u})\vec{u}$ and employing the identity $\hbar^2|\bn\tilde{\vec{s}}|^2/(8m^2\rho^2)=\hbar^2|\bn\vec{\psi}|^2/(2m^2\rho) - |\vec{u}|^2/2$ yield the momentum equation
\begin{equation}\label{eq:Du/Dt_1}
    \begin{aligned}
        \frac{\DD\vec{u}}{\DD t}
        =\ & - \frac{\hbar^2}{4m^2\rho}\bn\left(\frac{1}{\rho}|\bn\rho|^2 \right)
        + \frac{\hbar\nu'}{2m}\Delta\vec{u}
        + \frac{\hbar\nu'}{m\rho}|\bn\vec{\psi}|^2 \vec{u}
        \\
        &+ \bn\left(\frac{\hbar^2}{2m^2\rho}|\bn\vec{\psi}|^2 - \frac{|\vec{u}|^2}{2} - \frac{V_1}{m} + \frac{\hbar}{m\rho}\tilde{\vec{s}}\cdot\vec{f} \right)
        \\
        &+ \frac{\hbar^2}{4m^2\rho^3}\bn\rho\cdot\bigg[\bn(\rho\bn\rho) - \bn\bn\tilde{\vec{s}}\cdot\tilde{\vec{s}} + \frac{4m\nu'}{\hbar}\rho^2\bn\vec{u} \bigg]
        \\
        &+ \frac{\hbar\tilde{\vec{s}}\cdot\vec{f}}{m\rho^2}\bn\rho
        - \frac{\hbar}{m\rho}\bn\tilde{\vec{s}}\cdot\vec{f}
        + \frac{\hbar}{2m}(\bn\cdot\vec{u} + \vec{u}\cdot\bn\ln\rho)\bn\nu'.
    \end{aligned}
\end{equation}
To make Eq.~\eqref{eq:Du/Dt_1} consistent with the NSE, we set
\begin{equation}
    V_1 = \tilde{p} + \frac{\hbar^2}{2m\rho}|\bn\vec{\psi}|^2, \quad
    \nu' = \frac{2m\mu}{\hbar\rho},
\end{equation}
with a modified pressure
\begin{equation}\label{eq:SM_ptilde_compressible_viscous}
    \tilde{p} = \int\frac{m}{\rho}\dif\left(p - \frac{\hbar^2}{4m^2\rho}|\bn\rho|^2 - \frac{\mu}{3}\bn\cdot\vec{u} \right) - \frac{m}{2}|\vec{u}|^2 + \frac{\hbar}{\rho}\tilde{\vec{s}}\cdot\vec{f},
\end{equation}
where $p$ denotes the static pressure determined by the equation of state, and $\mu$ the viscosity.

Moreover, the vector $\vec{f}$ to be determined satisfies
\begin{equation}\label{eq:SM_condition_1}
    \frac{\tilde{\vec{s}}\cdot\vec{f}}{\rho}\bn\rho
    - \bn\tilde{\vec{s}}\cdot\vec{f}
    + \vec{\mathcal{S}}
    = \vec{0},
\end{equation}
with
\begin{equation}\label{eq:SM_compressible_viscous_S}
    \begin{aligned}
        \vec{\mathcal{S}}
        =\ & \frac{2m\mu}{\hbar\rho}|\bn\vec{\psi}|^2\vec{u}
        - \frac{m\mu}{\hbar}(\bn\cdot\vec{u} + \vec{u}\cdot\bn\ln\rho)\bn\ln\rho
        \\
        &+ \frac{\hbar}{4m\rho^2}\bn\rho\cdot\left[\bn(\rho\bn\rho) - \bn\bn\tilde{\vec{s}}\cdot\tilde{\vec{s}} + \frac{8m^2\mu}{\hbar^2}\rho\bn\vec{u} \right].
    \end{aligned}
\end{equation}
Introducing the coefficient matrix
\begin{equation}\label{eq:SM_A_compressible}
    \tensor{A} = \begin{bmatrix}
        \p_x\tilde{s}_1 - \tilde{s}_1\p_x\ln\rho & \p_x \tilde{s}_2 - \tilde{s}_2\p_x \ln\rho & \p_x\tilde{s}_3 - \tilde{s}_3\p_x\ln\rho \\
        \p_y\tilde{s}_1 - \tilde{s}_1\p_y\ln\rho & \p_y \tilde{s}_2 - \tilde{s}_2\p_y \ln\rho & \p_y\tilde{s}_3 - \tilde{s}_3\p_y\ln\rho \\
        \p_z\tilde{s}_1 - \tilde{s}_1\p_z\ln\rho & \p_z \tilde{s}_2 - \tilde{s}_2\p_z \ln\rho & \p_z\tilde{s}_3 - \tilde{s}_3\p_z\ln\rho
    \end{bmatrix},
\end{equation}
we recast Eq.~\eqref{eq:SM_condition_1} as a linear system $\tensor{A}\cdot\vec{f} = \vec{\mathcal{S}}$.
As
\begin{equation}
    \tilde{s}_j\p_i\tilde{s}_j - \tilde{s}_j\tilde{s}_j\p_i\ln\rho
    = \rho\p_i\rho - \rho^2\frac{\p_i\rho}{\rho}
    = 0,~ i = 1,2,3,
\end{equation}
the column rank of $\tensor{A}$ is not full and $\tensor{A}$ is non-invertible, so the solution to Eq.~\eqref{eq:SM_condition_1} is not unique.
We adopt the solution
\begin{equation}\label{eq:SM_f_compressible}
    \vec{f}
    = \tensor{A}^+\cdot\vec{\mathcal{S}},
\end{equation}
where $\tensor{A}^+=\lim_{\varepsilon\to 0}\tensor{A}\T(\tensor{A}\tensor{A}\T + \varepsilon\tensor{I})^{-1}$ is the Moore-Penrose pseudoinverse of $\tensor{A}$, which avoids the singularity for $|\tensor{A}|=0$ and minimizes $\|\vec{f}\|_2$.

In sum, we recast the NSE with constant $\mu$ in Eq.~\eqref{eq:compressible_NS} into a nonlinear SPE in the quaternion form
\begin{equation}\label{eq:Schrodinger_1}
    \begin{aligned}
        \vec{i}\hbar\frac{\p\vec{\psi}}{\p t}
        =\ & -\frac{\hbar^2}{2m}\left(1-\vec{i}\frac{2m\mu}{\hbar\rho}\right)\Delta\vec{\psi}
        + \left(\tilde{p} + \frac{\hbar^2}{2m\rho}|\bn\vec{\psi}|^2 \right)\vec{\psi}
        \\
        &+ \frac{\hbar\mu}{2}\frac{2|\bn\vec{\psi}|^2 - \Delta\rho}{\rho^2}\vec{i}\vec{\psi}
        - \frac{\hbar^2}{4m\rho}\vec{i\psi}\Delta\tilde{\vec{s}} + \hbar\vec{i\psi f}.
    \end{aligned}
\end{equation}

Finally, we transform the quaternion form in Eq.~\eqref{eq:Schrodinger_1} into the spin-1/2 form. 
Expanding the quaternion in Eq.~\eqref{eq:Schrodinger_1} yields
\begin{equation}\label{eq:supp_conversion_1}
\begin{aligned}
    &\vec{i}\hbar\frac{\p}{\p t}(a+\vec{i}b+\vec{j}c+\vec{k}d)
    \\
    =\ & -\frac{\hbar^2}{2m}\left(1-\vec{i}\frac{2m\mu}{\hbar\rho}\right)(\Delta a+\vec{i}\Delta b+\vec{j}\Delta c+\vec{k}\Delta d)
    \\
    &+ \left(\tilde{p} + \frac{\hbar^2}{2m\rho}|\bn\vec{\psi}|^2 \right)(a+\vec{i}b+\vec{j}c+\vec{k}d)
    \\
    &+ \frac{\hbar\mu}{2}\frac{2|\bn\vec{\psi}|^2 - \Delta\rho}{\rho^2}\vec{i}(a+\vec{i}b+\vec{j}c+\vec{k}d)
    \\
    &- \frac{\hbar^2}{4m\rho}(-b+\vec{i}a-\vec{j}d+\vec{k}c)(\vec{i}\Delta \tilde{s}_1 + \vec{j}\Delta \tilde{s}_2 + \vec{k}\Delta \tilde{s}_3)
    \\
    &+ \hbar(-b+\vec{i}a-\vec{j}d+\vec{k}c)(\vec{i}f_1 + \vec{j}f_2 + \vec{k}f_3).
\end{aligned}    
\end{equation}
All terms on the RHS of Eq.~\eqref{eq:supp_conversion_1} will be rearranged in terms of two complex-valued components.
The rearrangement for the first three terms is straightforward, and the treatment for the last two is similar.
For example, we expand the last term as
\begin{equation}\label{eq:supp_conversion_2}
\begin{aligned}
    & (-b+\vec{i}a-\vec{j}d+\vec{k}c)(\vec{i}f_1 + \vec{j}f_2 + \vec{k}f_3)
    \\
    =\ & -af_1 + df_2 - cf_3 + \vec{i}(-bf_1-df_3-cf_2) 
    \\
    &+ \vec{j}(-bf_2-af_3+cf_1) + \vec{k}(-bf_3+af_2+df_1)
    \\
    =\ & -f_1(a+\vec{i}b) - (f_3+\vec{i}f_2)(c+\vec{i}d) + f_1(\vec{j}c+\vec{k}d) 
    \\
    &- (f_3-\vec{i}f_2)(\vec{j}a+\vec{k}b).
\end{aligned}
\end{equation}

Thus, on the basis sets $\{1,\vec{i}\}$ and $\{\vec{j},\vec{k}\}$, Eq.~\eqref{eq:supp_conversion_2} can be recast into the standard spin-1/2 form as
\begin{equation}
    \begin{matrix}
        \{1,\vec{i}\}&: \\
        \{\vec{j},\vec{k}\}&:
    \end{matrix}
    \qquad
    \begin{bmatrix}
        -f_1 & -(f_3+\ii f_2) \\
        -(f_3-\ii f_2) & f_1
    \end{bmatrix}
    \begin{bmatrix}
        a+\ii b \\ c+\ii d
    \end{bmatrix}.
\end{equation}

By introducing the two-component wave function $\ket{\psi}=[\psi_+, \psi_-]\T$ with $\psi_+=a+\ii b$ and $\psi_-=c+\ii d$, we rewrite Eq.~\eqref{eq:Schrodinger_1} into
\begin{equation}\label{eq:SM_Schrodinger_2}
    \begin{aligned}
        &\ii\hbar\frac{\p}{\p t}\ket{\psi}
        \\
        =\ & \bigg[ \bigg(1-\ii\frac{2m\mu}{\hbar\rho}\bigg)\frac{\hat{\vec{p}}^2}{2m}
        + \tilde{p} + \frac{\hbar^2}{2m\rho}|\bn\psi|^2
        + \ii\frac{\hbar\mu}{2}\frac{2|\bn\psi|^2 - \Delta\rho}{\rho^2} \bigg]
        \ket{\psi}
        \\
        &+ \begin{bmatrix}
            \frac{\hbar^2\Delta \tilde{s}_1}{4m\rho} - \hbar f_1 & \frac{\hbar^2(\Delta \tilde{s}_3 + \ii\Delta \tilde{s}_2)}{4m\rho} - \hbar(f_3 + \ii f_2) \\
            \frac{\hbar^2(\Delta \tilde{s}_3 - \ii\Delta \tilde{s}_2)}{4m\rho} - \hbar(f_3 - \ii f_2) & -\frac{\hbar^2\Delta \tilde{s}_1}{4m\rho} + \hbar f_1
        \end{bmatrix}
        \ket{\psi},
    \end{aligned}
\end{equation}
where $\hat{\vec{p}}$ is the momentum operator, and the spin-vector components in terms of $\psi$ are
\begin{equation}
    \begin{cases}
        \tilde{s}_1 = |\psi_+|^2 - |\psi_-|^2, \\
        \tilde{s}_2 = \ii(\psi_+^*\psi_- - \psi_+\psi_-^*), \\
        \tilde{s}_3 = \psi_+^*\psi_- + \psi_+\psi_-^*.
    \end{cases}
\end{equation}
Then we introduce the Pauli vector
\begin{equation}\label{eq:Pauli_vector}
    \vec{\sigma} 
    = \sigma^j\vec{e}_j
    = \begin{bmatrix}
        0 & 1 \\ 1 & 0
    \end{bmatrix}\vec{e}_1
    + \begin{bmatrix}
        0 & -\ii \\ \ii & 0
    \end{bmatrix}\vec{e}_2
    + \begin{bmatrix}
        1 & 0 \\
        0 & -1
    \end{bmatrix}\vec{e}_3
\end{equation}
and replace the spin vector $\tilde{\vec{s}}$ by the average spin
\begin{equation}\label{eq:supp_average_spin}
    \vec{s} \equiv \langle\psi|\vec{\sigma}|\psi\rangle,
\end{equation}
where $\vec{s}$ in Eq.~\eqref{eq:supp_average_spin} can be obtained by reflecting the $y$-axis in $\mathbb{R}^3$ of $\tilde{\vec{s}}$ in the quaternionic form, i.e.
\begin{equation}
    s_1 = \tilde{s}_3, \quad s_2 = -\tilde{s}_2, \quad s_3 = \tilde{s}_1.
\end{equation}
Note that the vector $\vec{f}$ in $\vec{B}$ is calculated by replacing $\tilde{\vec{s}}$ by $\vec{s}$ in Eqs.~\eqref{eq:SM_compressible_viscous_S}, \eqref{eq:SM_A_compressible}, and \eqref{eq:SM_f_compressible}.
Finally, substituting Eq.~\eqref{eq:Pauli_vector} into Eq.~\eqref{eq:SM_Schrodinger_2} yields the SPE
\begin{equation}
    \begin{aligned}
        \ii\hbar\frac{\p}{\p t}\ket{\psi}
        =\ & \bigg[ \bigg(1-\ii\frac{2m\mu}{\hbar\rho}\bigg)\frac{\hat{\vec{p}}^2}{2m}
        + \tilde{p} + \frac{\hbar^2}{2m\rho}|\bn\psi|^2
        \\
        &+ \ii\frac{\hbar\mu}{2}\frac{2|\bn\psi|^2 - \Delta\rho}{\rho^2}\bigg] \tensor{I}\ket{\psi}
        + \vec{\sigma}\cdot\vec{B}\ket{\psi},
    \end{aligned}
\end{equation} 
i.e., Eq.~\eqref{eq:Schrodinger_2}.

%

\subsection{Derivation of the SPE for 3D incompressible flows}
We consider the incompressible flow with constant density $\rho=\rho_0$.
This flow is described by a special form of the SPE in Eq.~\eqref{eq:Schrodinger_2}.
To derive this SPE, Eq.~\eqref{eq:Du/Dt_1} can be simplified as
\begin{equation}\label{eq:Du/Dt_2}
    \begin{aligned}
        \frac{\DD\vec{u}}{\DD t}
        =\ & \bn\left(\frac{\hbar^2}{2m^2\rho_0}|\bn\vec{\psi}|^2 - \frac{|\vec{u}|^2}{2} - \frac{V_1}{m} \right)
        + \frac{\hbar\nu'}{2m}\Delta\vec{u}
        \\
        &+ \frac{\hbar}{m\rho_0}\bn\vec{f}\cdot\tilde{\vec{s}}
        + \frac{\hbar\nu'}{m\rho_0}|\bn\vec{\psi}|^2\vec{u}.
    \end{aligned}
\end{equation}
Then, we specify $\nu'=2m\nu/\hbar$ and enforce
\begin{equation}\label{eq:SM_f_condition_incomressible_1}
    \begin{aligned}
        &\bn\left(\frac{\hbar^2}{2m^2\rho_0}|\bn\vec{\psi}|^2 - \frac{|\vec{u}|^2}{2} - \frac{V_1}{m} \right) + \frac{\hbar}{m\rho_0}\bn\vec{f}\cdot\tilde{\vec{s}} + \frac{2\nu}{\rho_0}|\bn\vec{\psi}|^2\vec{u} 
        \\
        =\ & -\bn\frac{p}{\rho_0},
    \end{aligned}
\end{equation}
where $\nu\equiv\mu/\rho_0$ is the kinematic viscosity.
Applying the identity $\bn\vec{f}\cdot\tilde{\vec{s}} = \bn(\vec{f}\cdot\tilde{\vec{s}}) - \bn\tilde{\vec{s}}\cdot\vec{f}$ to Eq.~\eqref{eq:SM_f_condition_incomressible_1} yields
\begin{equation}\label{eq:SM_f_condition_incomressible_2}
    \frac{\hbar}{m\rho_0}\bn\tilde{\vec{s}}\cdot\vec{f} - \frac{2\nu}{\rho_0}|\bn\vec{\psi}|^2\vec{u}
    = \bn\phi_h,
\end{equation}
where
\begin{equation}\label{eq:SM_phih}
    \phi_h \equiv \frac{p}{\rho_0} + \frac{\hbar^2}{2m^2\rho_0}|\bn\vec{\psi}|^2 - \frac{|\vec{u}|^2}{2} - \frac{V_1}{m} + \frac{\hbar}{m\rho_0}\tilde{\vec{s}}\cdot\vec{f}
\end{equation}
is a real-valued function to be determined.

Projecting Eq.~\eqref{eq:SM_f_condition_incomressible_2} onto $\vec{\omega}$ and using the identity $\vec{\omega}\cdot\bn\tilde{\vec{s}}=\vec{0}$, we find
\begin{equation}\label{eq:SM_vor_dot_gradphih}
    \vec{\omega}\cdot\bn\phi_h = -\frac{2\nu}{\rho_0}|\bn\vec{\psi}|^2h,
\end{equation}
where $h\equiv \vec{u}\cdot\vec{\omega}$ denotes the helicity density.
We rewrite Eq.~\eqref{eq:SM_phih} into
\begin{equation}
    V_1 = \tilde{p} + \frac{\hbar^2}{2m\rho_0}|\bn\vec{\psi}|^2,
\end{equation}
with a modified pressure
\begin{equation}
    \tilde{p} = m\bigg(\frac{p}{\rho_0} - \frac{|\vec{u}|^2}{2} - \phi_h \bigg) + \frac{\hbar}{\rho_0}\tilde{\vec{s}}\cdot\vec{f}.
\end{equation}
Here, $\vec{f}$ is given by Eq.~\eqref{eq:SM_f_compressible} with
\begin{equation}\label{eq:SM_incompressible_A}
    \tensor{A}
    = \begin{bmatrix}
        \p_x \tilde{s}_1 & \p_x \tilde{s}_2 & \p_x \tilde{s}_3 \\
        \p_y \tilde{s}_1 & \p_y \tilde{s}_2 & \p_y \tilde{s}_3 \\
        \p_z \tilde{s}_1 & \p_z \tilde{s}_2 & \p_z \tilde{s}_3
    \end{bmatrix}
\end{equation}
and
\begin{equation}\label{eq:SM_S_incompressible_viscous}
    \vec{\mathcal{S}}
    = \frac{2m\nu}{\hbar}|\bn\vec{\psi}|^2\vec{u} - \frac{m\rho_0}{\hbar}\bn\phi_h.
\end{equation}

In sum, we reformulate the incompressible NSE in Eq.~\eqref{eq:incompressible_NS} into a nonlinear SPE in the quaternion form
\begin{equation}\label{eq:Schrodinger_3}
\begin{aligned}
    \vec{i}\hbar\frac{\p\vec{\psi}}{\p t}
    =\ & -\frac{\hbar^2}{2m}\left(1-\vec{i}\frac{2m\nu}{\hbar}\right)\Delta\vec{\psi} 
    - \frac{\hbar^2}{4m\rho_0}\vec{i\psi}\Delta\tilde{\vec{s}} + \hbar\vec{i\psi f}
    \\
    &+ \left(\tilde{p} + \frac{\hbar^2}{2m\rho_0}|\bn\vec{\psi}|^2 + \frac{\hbar\nu}{\rho_0}|\bn\vec{\psi}|^2\vec{i} \right)\vec{\psi}.
\end{aligned}
\end{equation}
Similarly, Eq.~\eqref{eq:Schrodinger_3} is re-expressed in terms of $\ket{\psi}$ as
\begin{equation}\label{eq:SM_Schrodinger_4}
\begin{aligned}
    \ii\hbar\frac{\p}{\p t}\ket{\psi}
    =\ & \bigg[\bigg(1 - \ii\frac{2m\nu}{\hbar} \bigg)\frac{\hat{\vec{p}}^2}{2m}
    + \tilde{p} + \frac{\hbar^2}{2m\rho_0}|\bn\psi|^2
    \\
    &+ \ii\frac{\hbar\nu}{\rho_0}|\bn\psi|^2 \bigg] \tensor{I}\ket{\psi}
    + \vec{\sigma}\cdot\vec{B}\ket{\psi},
\end{aligned}
\end{equation}
i.e., Eq.~\eqref{eq:Schrodinger_4}, with $\vec{B}=\hbar^2\Delta\vec{s}/(4m\rho_0)-\hbar\vec{f}$.

\section{Quantum representations of hydrodynamics: from simple to complex flows}\label{app:quantum_repre_steps}
%
As sketched in Fig.~\ref{fig:qm-fm_supp}, we summarize how to construct the SPE-NSE mapping, step by step, from simple to complex flows. 
With the same continuity equation in various flows, we only discuss the momentum equation below. 

\subsection{Potential flow}
The potential flow, with the vanishing vorticity $\bn\times\vec{u}=\vec{0}$, is governed by the momentum equation
\begin{equation}\label{eq:supp_momentum_potential}
    \frac{\p\vec{u}}{\p t} + \vec{u}\cdot\bn\vec{u} = -\bn\bigg(\frac{V}{m} - \frac{\hbar^2}{2m^2}\frac{\nabla^2\sqrt{\rho}}{\sqrt{\rho}} \bigg).
\end{equation}
Its quantum representation is given by the one-component Schrödinger equation 
\begin{equation}\label{eq:supp_Schr_potential}
    \ii\hbar\frac{\p\psi}{\p t} = \bigg(\frac{\hat{\vec{p}}^2}{2m} + V \bigg)\psi.
\end{equation}
With $V=V(\vec{x})$, Eq.~\eqref{eq:supp_Schr_potential} is linear, and the RHS of Eq.~\eqref{eq:supp_momentum_potential} represents an artificial conservative force.
Setting a nonlinear potential $V=\hbar^2\nabla^2|\psi|/(2m|\psi|)$ renders Eq.~\eqref{eq:supp_Schr_potential} nonlinear, and transforms Eq.~\eqref{eq:supp_momentum_potential} into the Burgers equation with vanishing vorticity.

\subsection{Schrödinger flow}
Due to the very limited application of the potential flow, it is essential to introduce finite vorticity into the hydrodynamic formulation of the Schrödinger equation by introducing a two-component wave function $\ket{\psi}=[\psi_+,\psi_-]\T$. 

The resultant Schrödinger flow~\cite{Meng2023_Quantum} is governed by the momentum equation
\begin{equation}\label{eq:supp_momentum_Schr}
    \frac{\p\vec{u}}{\p t} + \vec{u}\cdot\bn\vec{u}
    = -\bn \bigg(\frac{V}{m} - \frac{\hbar^2|\bn\vec{s}|^2}{8m^2\rho^2} \bigg) -\frac{\hbar}{m\rho}\bn(\vec{\zeta}\cdot\vec{s}) + \frac{\hbar}{m\rho}\bn\vec{s}\cdot\vec{\zeta},
\end{equation}
with $\vec{\zeta}=-\bn\cdot(\bn\vec{s}/\rho)/4$. 
Its quantum representation is given by the two-component Schrödinger equation
\begin{equation}
    \ii\hbar\frac{\p}{\p t}\ket{\psi} = \bigg(\frac{\hat{\vec{p}}^2}{2m} + V \bigg)\ket{\psi}.
\end{equation}
The terms on the RHS of Eq.~\eqref{eq:supp_momentum_Schr} represent the artificial conservative body force, the pressure from a specific equation of state, and the Landau-Lifshitz force. 
It has been shown that the incompressible Schrödinger flow with $V=\hbar^2|\bn\vec{s}|^2/(8m\rho^2)$ can resemble a turbulent flow consisting of tangled vortex tubes with the $-5/3$ scaling of energy spectrum~\cite{Meng2023_Quantum}.

\subsection{Euler flow}
To obtain the proper pressure and eliminate the artificial body force in the Schrödinger flow, a nonlinear potential and a spin-coupling term (i.e., the Stern-Gerlach term) are introduced into the two-component Schrödinger equation. This modification leads  the Euler flow~\cite{Meng2024_Lagrangian}, governed by the momentum equation in Eq.~\eqref{eq:Euler_Eq}.
Its quantum representation is given by the SPE in Eq.~\eqref{eq:Schrodinger_5}, with the modified pressure $\tilde{p}$ in Eq.~\eqref{eq:SM_tildep_compressible_inviscid} and the spin-coupled vector $\vec{B}=\hbar^2\Delta\vec{s}/(4m\rho) - \hbar\vec{f}$.


In particular, this quantum formulation provides a distinct Lagrangian viewpoint for studying vortex dynamics in an incompressible inviscid flow~\cite{Meng2024_Lagrangian}. 

\subsection{Navier-Stokes flow}
To extend the quantum representation from inviscid flow to viscous flow, the viscous dissipation is incorporated via the imaginary diffusion of the wave function~\cite{Salasnich2024_Quantum}. 

The resultant Navier-Stokes flow is governed by the momentum equation in Eq.~\eqref{eq:compressible_NS}.
Its quantum representation is given by the SPE in Eq.~\eqref{eq:Schrodinger_2} with the modified pressure $\tilde{p}$ in Eq.~\eqref{eq:SM_ptilde_compressible_viscous} and the spin-coupled vector $\vec{B}=\hbar^2\Delta\vec{s}/(4m\rho) - \hbar\vec{f}$. 

The imaginary diffusion $-\ii \mu\hat{\vec{p}}^2/(\hbar\rho)\tensor{I}\ket{\psi}$ in Eq.~\eqref{eq:Schrodinger_2} corresponds to the dissipative term $\mu\Delta\vec{u}/\rho$ in Eq.~\eqref{eq:compressible_NS}, and the imaginary potential $\ii\hbar\mu(2|\bn\psi|^2-\Delta\rho)/(2\rho^2)$ introduces a compressive effect, thereby preserving probability.

\section{Schr\"odinger-Pauli formulation of the 1D Burgers equation}\label{app:SPE_Burgers}
\subsection{Derivation of the SPE for a 1D compressible flow}
We derive the SPE equivalent to the Burgers equation
\begin{equation}\label{eq:Burgers}
    u_t + uu_x = \nu u_{xx}.
\end{equation}
Setting $\nu'=2m\nu/\hbar$, the 1D form of Eq.~\eqref{eq:Du/Dt_1} reduces to
\begin{equation}
    \begin{aligned}
        u_t + uu_x
        =\ & \nu u_{xx}
        + \frac{\hbar\rho_x}{m\rho^2}\bigg[\tilde{\vec{s}}\cdot\vec{f} + \frac{\hbar}{4m\rho}\left(\rho_x^2 - 2\rho\rho_{xx} + |\tilde{\vec{s}}_x|^2\right) \bigg]
        \\
        &- \frac{\hbar}{m\rho}\tilde{\vec{s}}_x\cdot\vec{f}
        + \frac{2\nu}{\rho}\left(\rho_xu_x + |\vec{\psi}_x|^2u \right)
        \\
        &+ \frac{\p}{\p x}\bigg(\frac{\hbar^2}{8m^2\rho^2}|\tilde{\vec{s}}_x|^2 - \frac{V_1}{m} + \frac{\hbar}{m\rho}\tilde{\vec{s}}\cdot\vec{f} \bigg).
    \end{aligned}
\end{equation}
Subsequently, we enforce
\begin{equation}\label{eq:f_constraints_Burgers}
    \begin{dcases}
        \tilde{\vec{s}}\cdot\vec{f} + \frac{\hbar}{4m\rho}\left(\rho_x^2 - 2\rho\rho_{xx} + |\tilde{\vec{s}}_x|^2 \right) = 0, \\
        - \frac{\hbar}{m\rho}\tilde{\vec{s}}_x\cdot\vec{f}
        + \frac{2\nu}{\rho}\left(\rho_xu_x + |\vec{\psi}_x|^2u\right) = 0, \\
        \frac{\hbar^2}{8m^2\rho^2}|\tilde{\vec{s}}_x|^2 - \frac{V_1}{m} + \frac{\hbar}{m\rho}\tilde{\vec{s}}\cdot\vec{f} = 0.
    \end{dcases}
\end{equation}
The solutions to Eq.~\eqref{eq:f_constraints_Burgers} are
\begin{equation}\label{eq:SM_Burgers_f}
    \vec{f} = \lambda_1\tilde{\vec{s}} + \lambda_2\tilde{\vec{s}}_x, \quad
    V_1 = -\frac{\hbar^2}{4m\rho^2}\bigg(\rho_x^2 - 2\rho\rho_{xx} + \frac{1}{2}|\tilde{\vec{s}}_x|^2 \bigg)
\end{equation}
with
\begin{equation}\label{eq:SM_lambda12}
    \begin{bmatrix}
        \lambda_1 \\ \lambda_2
    \end{bmatrix}
    = \frac{1}{\rho^2(|\tilde{\vec{s}}_x|^2 - \rho_x^2)}
    \begin{bmatrix}
        |\tilde{\vec{s}}_x|^2 & -\rho\rho_x \\
        -\rho\rho_x & \rho^2
    \end{bmatrix}
    \begin{bmatrix}
        -\frac{\hbar\left(\rho_x^2 - 2\rho\rho_{xx} + |\tilde{\vec{s}}_x|^2 \right)}{4m\rho} \\
        \frac{2m\nu\left(\rho_xu_x + |\vec{\psi}_x|^2u \right)}{\hbar}
    \end{bmatrix}.
\end{equation}
Hence, the equivalent SPE for the Burgers equation~\eqref{eq:Burgers} reads
\begin{equation}\label{eq:SM_SPE_Burgers}
    \begin{aligned}
        \ii\hbar\frac{\p}{\p t}\ket{\psi}
        =\ & \bigg[\bigg(1 - \ii\frac{2m\nu}{\hbar} \bigg)\frac{\hat{\vec{p}}^2}{2m} -\frac{\hbar^2}{4m\rho^2}\bigg(\rho_x^2 - 2\rho\rho_{xx} + \frac{1}{2}|\vec{s}_x|^2 \bigg)
        \\
        &+ \ii\frac{\hbar\nu(2|\psi_x|^2 - \rho_{xx})}{2\rho} \bigg]\tensor{I}\ket{\psi}
        + \vec{\sigma}\cdot\vec{B}\ket{\psi},
    \end{aligned}
\end{equation}
with $\vec{B}=\hbar^2\vec{s}_{xx}/(4m\rho) - \hbar\vec{f}$.

\subsection{Numerical algorithm}
We develop a numerical algorithm for solving Eq.~\eqref{eq:SM_SPE_Burgers}. Here $m=1$ and $\hbar=1$ are set without loss of generality.
The solution domain $x\in\mathcal{D}$ is discretized on uniform grid points $x_j,~j=1,2,\cdots,N$ with mesh spacing $\dx$.
The second-order central difference and the first-order backward difference are employed for the spatial discretization and time marching, respectively.
The discretized Eq.~\eqref{eq:SM_SPE_Burgers} reads
\begin{equation}\label{eq:SM_SPE_Burgers_1}
    \begin{cases}
    \begin{aligned}
        \ii\frac{\psi_{+,j}^{n+1} - \psi_{+,j}^n}{\dt}
        =\ & \bigg(\ii\nu - \frac{1}{2}\bigg)\frac{\psi_{+,j+1}^{n+1} - 2\psi_{+,j}^{n+1} + \psi_{+,j-1}^{n+1}}{\dx^2}
        \\
        &+ (V_{j}^n + \ii W_{j}^n)\psi_{+,j}^n
        + P_j^n\psi_{+,j}^n + Q_j^n\psi_{-,j}^n, \\
        \ii\frac{\psi_{-,j}^{n+1} - \psi_{-,j}^n}{\dt}
        =\ & \bigg(\ii\nu - \frac{1}{2}\bigg)\frac{\psi_{-,j+1}^{n+1} - 2\psi_{-,j}^{n+1} + \psi_{-,j-1}^{n+1}}{\dx^2}
        \\
        &+ (V_{j}^n + \ii W_{j}^n)\psi_{-,j}^n
        + (Q_j^n)^*\psi_{+,j}^n - P_j^n\psi_{-,j}^n,
    \end{aligned}
    \end{cases}
\end{equation}
with the time stepping $\dt$ and potentials
\begin{equation}\label{eq:SM_Burgers_V_W}
    V = -\frac{\hbar^2}{4m\rho^2}\bigg(\rho_x^2 - 2\rho\rho_{xx} + \frac{1}{2}|\vec{s}_x|^2 \bigg), \quad
    W = \frac{\nu(2|\vec{\psi}_x|^2 - \rho_{xx})}{2\rho},
\end{equation}
and
\begin{equation}\label{eq:SM_Burgers_P_Q}
    P = \frac{\Delta s_1}{4\rho} - f_1, \quad
    Q = \frac{\Delta s_3 + \ii\Delta s_2}{4\rho} - (f_3 + \ii f_2).
\end{equation}

We impose the Neumann boundary conditions on the wave function at the boundary as
\begin{equation}\label{eq:SM_SPE_Burgers_boundary}
    \frac{\psi_{\pm,2} - \psi_{\pm,0}}{2\dx} = 0, \quad
    \frac{\psi_{\pm,N+1} - \psi_{\pm,N-1}}{2\dx} = 0.
\end{equation}
Then, we obtain two linear systems
\begin{equation}\label{eq:SM_Burgers_linear_system}
    \begin{cases}
    \begin{aligned}
        &\alpha\psi_{+,j+1}^{n+1} - (2\alpha+1)\psi_{+,j}^{n+1} + \alpha\psi_{+,j-1}^{n+1}
        \\
        =\ & \ii\dt\left[(V_{j}^n + P_j^n + \ii W_{j}^n)\psi_{+,j}^n + Q_j^n\psi_{-,j}^n\right] - \psi_{+,j}^n, \\
        &\alpha\psi_{-,j+1}^{n+1} - (2\alpha+1)\psi_{-,j}^{n+1} + \alpha\psi_{-,j-1}^{n+1}
        \\
        =\ & \ii\dt\left[(V_{j}^n - P_j^n + \ii W_{j}^n)\psi_{-,j}^n + (Q_j^n)^*\psi_{+,j}^n\right] - \psi_{-,j}^n
    \end{aligned}    
    \end{cases}
\end{equation}
from Eqs.~\eqref{eq:SM_SPE_Burgers_1} and \eqref{eq:SM_SPE_Burgers_boundary} with $\alpha=(\nu+\ii/2)\dt/\dx^2$.
The tridiagonal matrix in Eq.~\eqref{eq:SM_Burgers_linear_system} is strictly diagonally dominant due to
\begin{equation}
    |2\alpha+1|
    = \frac{\dt}{\dx^2}\sqrt{\bigg(2\nu + \frac{\dx^2}{\dt} \bigg)^2 + 1}
    > \frac{\dt}{\dx^2}\sqrt{(2\nu )^2 + 1}
    = |2\alpha|,
\end{equation}
so that Eq.~\eqref{eq:SM_Burgers_linear_system} can be numerically solved efficiently by the catch-up method.
The numerical method is presented in Algorithm~\ref{ag:Burgers}.

\begin{algorithm}\label{ag:Burgers}
    \caption{Time iteration of the SPE equivalent to the Burgers equation.}
    \KwIn{$\ket{\psi(t)}$, $\nu$, $\dx$, $\dt$, $N$}
    \KwOut{$\ket{\psi(t+\dt)}$}
    $\psi_{\pm,j}^n \leftarrow \ket{\psi(t)}$\;
    \For{$j \leftarrow 1$ \KwTo $N$}{
        Calculate the fluid density $\rho_j^n$ by Eq.~\eqref{eq:rho} and the velocity $u_j^n$ by Eq.~\eqref{eq:vel}\;
        Calculate the spin vector $\vec{s}_j^n$ by Eq.~\eqref{eq:spin_vector}\;
        Calculate $(\rho_{x})_j^n$, $(\rho_{xx})_j^n$, $(u_x)_j^n$, $(\vec{s}_x)_j^n$, and $(\psi_{\pm,x})_j^n$ using second-order central difference\;
        Calculate $(\lambda_1)_j^n$ and $(\lambda_2)_j^n$ by Eq.~\eqref{eq:SM_lambda12}\;
        Calculate $\vec{f}_j^n$ by Eq.~\eqref{eq:SM_Burgers_f}\;
        Calculate the potentials $V_j^n$, $W_j^n$, $P_j^n$, and $Q_j^n$ by Eqs.~\eqref{eq:SM_Burgers_V_W} and \eqref{eq:SM_Burgers_P_Q}\;
        Obtain $\psi_{\pm,j}^{n+1}$ by solving the linear systems in Eq.~\eqref{eq:SM_Burgers_linear_system} using catch-up method\;
    }
    $\ket{\psi(t+\dt)} \leftarrow \psi_{\pm,j}^{n+1}$.
\end{algorithm}

\let\oldaddcontentsline\addcontentsline
\renewcommand{\addcontentsline}[3]{}
\bibliographystyle{modified-apsrev4-2.bst}

\begin{thebibliography}{63}%
\makeatletter
\providecommand \@ifxundefined [1]{%
 \@ifx{#1\undefined}
}%
\providecommand \@ifnum [1]{%
 \ifnum #1\expandafter \@firstoftwo
 \else \expandafter \@secondoftwo
 \fi
}%
\providecommand \@ifx [1]{%
 \ifx #1\expandafter \@firstoftwo
 \else \expandafter \@secondoftwo
 \fi
}%
\providecommand \natexlab [1]{#1}%
\providecommand \enquote  [1]{``#1''}%
\providecommand \bibnamefont  [1]{#1}%
\providecommand \bibfnamefont [1]{#1}%
\providecommand \citenamefont [1]{#1}%
\providecommand \href@noop [0]{\@secondoftwo}%
\providecommand \href [0]{\begingroup \@sanitize@url \@href}%
\providecommand \@href[1]{\@@startlink{#1}\@@href}%
\providecommand \@@href[1]{\endgroup#1\@@endlink}%
\providecommand \@sanitize@url [0]{\catcode `\\12\catcode `\$12\catcode `\&12\catcode `\#12\catcode `\^12\catcode `\_12\catcode `\%12\relax}%
\providecommand \@@startlink[1]{}%
\providecommand \@@endlink[0]{}%
\providecommand \url  [0]{\begingroup\@sanitize@url \@url }%
\providecommand \@url [1]{\endgroup\@href {#1}{\urlprefix }}%
\providecommand \urlprefix  [0]{URL }%
\providecommand \Eprint [0]{\href }%
\providecommand \doibase [0]{https://doi.org/}%
\providecommand \selectlanguage [0]{\@gobble}%
\providecommand \bibinfo  [0]{\@secondoftwo}%
\providecommand \bibfield  [0]{\@secondoftwo}%
\providecommand \translation [1]{[#1]}%
\providecommand \BibitemOpen [0]{}%
\providecommand \bibitemStop [0]{}%
\providecommand \bibitemNoStop [0]{.\EOS\space}%
\providecommand \EOS [0]{\spacefactor3000\relax}%
\providecommand \BibitemShut  [1]{\csname bibitem#1\endcsname}%
\let\auto@bib@innerbib\@empty
\bibitem [{\citenamefont {Gilet}\ and\ \citenamefont {Bush}(2009)}]{Gilet2009_Chaotic}%
  \BibitemOpen
  \bibfield  {author} {\bibinfo {author} {\bibfnamefont {T.}~\bibnamefont {Gilet}}\ and\ \bibinfo {author} {\bibfnamefont {J.~W.~M.}\ \bibnamefont {Bush}},\ }\bibfield  {title} {\bibinfo {title} {{\color{Black}{Chaotic bouncing of a droplet on a soap film}}},\ }\href {https://doi.org/10.1103/PhysRevLett.102.014501} {\bibfield  {journal} {\bibinfo  {journal} {Phys. Rev. Lett.}\ }\textbf {\bibinfo {volume} {102}},\ \bibinfo {pages} {014501} (\bibinfo {year} {2009})}\BibitemShut {NoStop}%
\bibitem [{\citenamefont {Bush}\ and\ \citenamefont {Oza}(2020)}]{Bush2020_Hydrodynamic}%
  \BibitemOpen
  \bibfield  {author} {\bibinfo {author} {\bibfnamefont {J.~W.~M.}\ \bibnamefont {Bush}}\ and\ \bibinfo {author} {\bibfnamefont {A.~U.}\ \bibnamefont {Oza}},\ }\bibfield  {title} {\bibinfo {title} {{\color{Black}{Hydrodynamic quantum analogs}}},\ }\href {https://doi.org/10.1088/1361-6633/abc22c} {\bibfield  {journal} {\bibinfo  {journal} {Rep. Prog. Phys.}\ }\textbf {\bibinfo {volume} {84}},\ \bibinfo {pages} {017001} (\bibinfo {year} {2020})}\BibitemShut {NoStop}%
\bibitem [{\citenamefont {Sáenz}\ \emph {et~al.}(2021)\citenamefont {Sáenz}, \citenamefont {Pucci}, \citenamefont {Turton}, \citenamefont {Goujon}, \citenamefont {Rosales}, \citenamefont {Dunkel},\ and\ \citenamefont {Bush}}]{Saenz2021_Emergent}%
  \BibitemOpen
  \bibfield  {author} {\bibinfo {author} {\bibfnamefont {P.~J.}\ \bibnamefont {Sáenz}}, \bibinfo {author} {\bibfnamefont {G.}~\bibnamefont {Pucci}}, \bibinfo {author} {\bibfnamefont {S.~E.}\ \bibnamefont {Turton}}, \bibinfo {author} {\bibfnamefont {A.}~\bibnamefont {Goujon}}, \bibinfo {author} {\bibfnamefont {R.~R.}\ \bibnamefont {Rosales}}, \bibinfo {author} {\bibfnamefont {J.}~\bibnamefont {Dunkel}},\ and\ \bibinfo {author} {\bibfnamefont {J.~W.~M.}\ \bibnamefont {Bush}},\ }\bibfield  {title} {\bibinfo {title} {{\color{Black}{Emergent order in hydrodynamic spin lattices}}},\ }\href {https://doi.org/10.1038/s41586-021-03682-1} {\bibfield  {journal} {\bibinfo  {journal} {Nature}\ }\textbf {\bibinfo {volume} {596}},\ \bibinfo {pages} {58} (\bibinfo {year} {2021})}\BibitemShut {NoStop}%
\bibitem [{\citenamefont {Kardar}\ and\ \citenamefont {Golestanian}(1999)}]{Kardar1999_The}%
  \BibitemOpen
  \bibfield  {author} {\bibinfo {author} {\bibfnamefont {M.}~\bibnamefont {Kardar}}\ and\ \bibinfo {author} {\bibfnamefont {R.}~\bibnamefont {Golestanian}},\ }\bibfield  {title} {\bibinfo {title} {{\color{Black}{The ``friction'' of vacuum, and other fluctuation-induced forces}}},\ }\href {https://doi.org/10.1103/RevModPhys.71.1233} {\bibfield  {journal} {\bibinfo  {journal} {Rev. Mod. Phys.}\ }\textbf {\bibinfo {volume} {71}},\ \bibinfo {pages} {1233} (\bibinfo {year} {1999})}\BibitemShut {NoStop}%
\bibitem [{\citenamefont {Davoodianidalik}\ \emph {et~al.}(2022)\citenamefont {Davoodianidalik}, \citenamefont {Punzmann}, \citenamefont {Kellay}, \citenamefont {Xia}, \citenamefont {Shats},\ and\ \citenamefont {Francois}}]{Davoodianidalik2022_Fluctuation}%
  \BibitemOpen
  \bibfield  {author} {\bibinfo {author} {\bibfnamefont {M.}~\bibnamefont {Davoodianidalik}}, \bibinfo {author} {\bibfnamefont {H.}~\bibnamefont {Punzmann}}, \bibinfo {author} {\bibfnamefont {H.}~\bibnamefont {Kellay}}, \bibinfo {author} {\bibfnamefont {H.}~\bibnamefont {Xia}}, \bibinfo {author} {\bibfnamefont {M.}~\bibnamefont {Shats}},\ and\ \bibinfo {author} {\bibfnamefont {N.}~\bibnamefont {Francois}},\ }\bibfield  {title} {\bibinfo {title} {{\color{Black}{Fluctuation-induced interaction in turbulent flows}}},\ }\href {https://doi.org/10.1103/PhysRevLett.128.024503} {\bibfield  {journal} {\bibinfo  {journal} {Phys. Rev. Lett.}\ }\textbf {\bibinfo {volume} {128}},\ \bibinfo {pages} {024503} (\bibinfo {year} {2022})}\BibitemShut {NoStop}%
\bibitem [{\citenamefont {Mendonça}\ \emph {et~al.}(2001)\citenamefont {Mendonça}, \citenamefont {Bingham}, \citenamefont {Shukla},\ and\ \citenamefont {Resendes}}]{Mendonca2001_Casimir}%
  \BibitemOpen
  \bibfield  {author} {\bibinfo {author} {\bibfnamefont {J.~T.}\ \bibnamefont {Mendonça}}, \bibinfo {author} {\bibfnamefont {R.}~\bibnamefont {Bingham}}, \bibinfo {author} {\bibfnamefont {P.~K.}\ \bibnamefont {Shukla}},\ and\ \bibinfo {author} {\bibfnamefont {D.}~\bibnamefont {Resendes}},\ }\bibfield  {title} {\bibinfo {title} {{\color{Black}{Casimir effect in a turbulent plasma}}},\ }\href {https://doi.org/10.1016/S0375-9601(01)00614-4} {\bibfield  {journal} {\bibinfo  {journal} {Phys. Lett. A}\ }\textbf {\bibinfo {volume} {289}},\ \bibinfo {pages} {233} (\bibinfo {year} {2001})}\BibitemShut {NoStop}%
\bibitem [{\citenamefont {Agam}\ \emph {et~al.}(2002)\citenamefont {Agam}, \citenamefont {Bettelheim}, \citenamefont {Wiegmann},\ and\ \citenamefont {Zabrodin}}]{Agam2002_Viscous}%
  \BibitemOpen
  \bibfield  {author} {\bibinfo {author} {\bibfnamefont {O.}~\bibnamefont {Agam}}, \bibinfo {author} {\bibfnamefont {E.}~\bibnamefont {Bettelheim}}, \bibinfo {author} {\bibfnamefont {P.}~\bibnamefont {Wiegmann}},\ and\ \bibinfo {author} {\bibfnamefont {A.}~\bibnamefont {Zabrodin}},\ }\bibfield  {title} {\bibinfo {title} {{\color{Black}{Viscous fingering and the shape of an electronic droplet in the quantum Hall regime}}},\ }\href {https://doi.org/10.1103/PhysRevLett.88.236801} {\bibfield  {journal} {\bibinfo  {journal} {Phys. Rev. Lett.}\ }\textbf {\bibinfo {volume} {88}},\ \bibinfo {pages} {236801} (\bibinfo {year} {2002})}\BibitemShut {NoStop}%
\bibitem [{\citenamefont {Zu}\ \emph {et~al.}(2021)\citenamefont {Zu}, \citenamefont {Machado}, \citenamefont {Ye}, \citenamefont {Choi}, \citenamefont {Kobrin}, \citenamefont {Mittiga}, \citenamefont {Hsieh}, \citenamefont {Bhattacharyya}, \citenamefont {Markham}, \citenamefont {Twitchen}, \citenamefont {Jarmola}, \citenamefont {Budker}, \citenamefont {Laumann}, \citenamefont {Moore},\ and\ \citenamefont {Yao}}]{Zu2021_Emergent}%
  \BibitemOpen
  \bibfield  {author} {\bibinfo {author} {\bibfnamefont {C.}~\bibnamefont {Zu}}, \bibinfo {author} {\bibfnamefont {F.}~\bibnamefont {Machado}}, \bibinfo {author} {\bibfnamefont {B.}~\bibnamefont {Ye}}, \bibinfo {author} {\bibfnamefont {S.}~\bibnamefont {Choi}}, \bibinfo {author} {\bibfnamefont {B.}~\bibnamefont {Kobrin}}, \bibinfo {author} {\bibfnamefont {T.}~\bibnamefont {Mittiga}}, \bibinfo {author} {\bibfnamefont {S.}~\bibnamefont {Hsieh}}, \bibinfo {author} {\bibfnamefont {P.}~\bibnamefont {Bhattacharyya}}, \bibinfo {author} {\bibfnamefont {M.}~\bibnamefont {Markham}}, \bibinfo {author} {\bibfnamefont {D.}~\bibnamefont {Twitchen}}, \emph {et~al.},\ }\bibfield  {title} {\bibinfo {title} {{\color{Black}{Emergent hydrodynamics in a strongly interacting dipolar spin ensemble}}},\ }\href {https://doi.org/10.1038/s41586-021-03763-1} {\bibfield  {journal} {\bibinfo  {journal} {Nature}\ }\textbf {\bibinfo {volume} {597}},\ \bibinfo {pages} {45} (\bibinfo {year} {2021})}\BibitemShut {NoStop}%
\bibitem [{\citenamefont {Joshi}\ \emph {et~al.}(2022)\citenamefont {Joshi}, \citenamefont {Kranzl}, \citenamefont {Schuckert}, \citenamefont {Lovas}, \citenamefont {Maier}, \citenamefont {Blatt}, \citenamefont {Knap},\ and\ \citenamefont {Roos}}]{Joshi2022_Observing}%
  \BibitemOpen
  \bibfield  {author} {\bibinfo {author} {\bibfnamefont {M.~K.}\ \bibnamefont {Joshi}}, \bibinfo {author} {\bibfnamefont {F.}~\bibnamefont {Kranzl}}, \bibinfo {author} {\bibfnamefont {A.}~\bibnamefont {Schuckert}}, \bibinfo {author} {\bibfnamefont {I.}~\bibnamefont {Lovas}}, \bibinfo {author} {\bibfnamefont {C.}~\bibnamefont {Maier}}, \bibinfo {author} {\bibfnamefont {R.}~\bibnamefont {Blatt}}, \bibinfo {author} {\bibfnamefont {M.}~\bibnamefont {Knap}},\ and\ \bibinfo {author} {\bibfnamefont {C.~F.}\ \bibnamefont {Roos}},\ }\bibfield  {title} {\bibinfo {title} {{\color{Black}{Observing emergent hydrodynamics in a long-range quantum magnet}}},\ }\href {https://doi.org/10.1126/science.abk2400} {\bibfield  {journal} {\bibinfo  {journal} {Science}\ }\textbf {\bibinfo {volume} {376}},\ \bibinfo {pages} {720} (\bibinfo {year} {2022})}\BibitemShut {NoStop}%
\bibitem [{\citenamefont {Gross}(1961)}]{Gross1961_Structure}%
  \BibitemOpen
  \bibfield  {author} {\bibinfo {author} {\bibfnamefont {E.~P.}\ \bibnamefont {Gross}},\ }\bibfield  {title} {\bibinfo {title} {{\color{Black}{Structure of a quantized vortex in boson systems}}},\ }\href {https://doi.org/10.1007/BF02731494} {\bibfield  {journal} {\bibinfo  {journal} {Il Nuovo Cimento}\ }\textbf {\bibinfo {volume} {20}},\ \bibinfo {pages} {454} (\bibinfo {year} {1961})}\BibitemShut {NoStop}%
\bibitem [{\citenamefont {Pitaevskii}(1961)}]{Pitaevskii1961_Vortex}%
  \BibitemOpen
  \bibfield  {author} {\bibinfo {author} {\bibfnamefont {L.~P.}\ \bibnamefont {Pitaevskii}},\ }\bibfield  {title} {\bibinfo {title} {{\color{Black}{Vortex lines in an imperfect Bose gas}}},\ }\href {http://www.jetp.ras.ru/cgi-bin/e/index/e/13/2/p451?a=list} {\bibfield  {journal} {\bibinfo  {journal} {Sov. Phys. JETP.}\ }\textbf {\bibinfo {volume} {13}},\ \bibinfo {pages} {451} (\bibinfo {year} {1961})}\BibitemShut {NoStop}%
\bibitem [{\citenamefont {Marklund}\ and\ \citenamefont {Brodin}(2007)}]{Marklund2007_Dynamics}%
  \BibitemOpen
  \bibfield  {author} {\bibinfo {author} {\bibfnamefont {M.}~\bibnamefont {Marklund}}\ and\ \bibinfo {author} {\bibfnamefont {G.}~\bibnamefont {Brodin}},\ }\bibfield  {title} {\bibinfo {title} {{\color{Black}{Dynamics of spin-1/2 quantum plasmas}}},\ }\href {https://doi.org/10.1103/PhysRevLett.98.025001} {\bibfield  {journal} {\bibinfo  {journal} {Phys. Rev. Lett.}\ }\textbf {\bibinfo {volume} {98}},\ \bibinfo {pages} {025001} (\bibinfo {year} {2007})}\BibitemShut {NoStop}%
\bibitem [{\citenamefont {Love}\ and\ \citenamefont {Boghosian}(2004)}]{Love2004_Quaternionic}%
  \BibitemOpen
  \bibfield  {author} {\bibinfo {author} {\bibfnamefont {P.~J.}\ \bibnamefont {Love}}\ and\ \bibinfo {author} {\bibfnamefont {B.~M.}\ \bibnamefont {Boghosian}},\ }\bibfield  {title} {\bibinfo {title} {{\color{Black}{Quaternionic Madelung transformation and non-Abelian fluid dynamics}}},\ }\href {https://doi.org/10.1016/j.physa.2003.09.055} {\bibfield  {journal} {\bibinfo  {journal} {Physica A}\ }\textbf {\bibinfo {volume} {332}},\ \bibinfo {pages} {47} (\bibinfo {year} {2004})}\BibitemShut {NoStop}%
\bibitem [{\citenamefont {Hasimoto}(1972)}]{Hasimoto1972_A}%
  \BibitemOpen
  \bibfield  {author} {\bibinfo {author} {\bibfnamefont {H.}~\bibnamefont {Hasimoto}},\ }\bibfield  {title} {\bibinfo {title} {{\color{Black}{A soliton on a vortex filament}}},\ }\href {https://doi.org/10.1017/s0022112072002307} {\bibfield  {journal} {\bibinfo  {journal} {J. Fluid Mech.}\ }\textbf {\bibinfo {volume} {51}},\ \bibinfo {pages} {477} (\bibinfo {year} {1972})}\BibitemShut {NoStop}%
\bibitem [{\citenamefont {Sone}\ and\ \citenamefont {Ashida}(2019)}]{Sone2019_Anomalous}%
  \BibitemOpen
  \bibfield  {author} {\bibinfo {author} {\bibfnamefont {K.}~\bibnamefont {Sone}}\ and\ \bibinfo {author} {\bibfnamefont {Y.}~\bibnamefont {Ashida}},\ }\bibfield  {title} {\bibinfo {title} {{\color{Black}{Anomalous topological active matter}}},\ }\href {https://doi.org/10.1103/PhysRevLett.123.205502} {\bibfield  {journal} {\bibinfo  {journal} {Phys. Rev. Lett.}\ }\textbf {\bibinfo {volume} {123}},\ \bibinfo {pages} {205502} (\bibinfo {year} {2019})}\BibitemShut {NoStop}%
\bibitem [{\citenamefont {Madelung}(1927)}]{Madelung1927_Quantentheorie}%
  \BibitemOpen
  \bibfield  {author} {\bibinfo {author} {\bibfnamefont {E.}~\bibnamefont {Madelung}},\ }\bibfield  {title} {\bibinfo {title} {{\color{Black}{Quantentheorie in hydrodynamischer form}}},\ }\href {https://doi.org/10.1007/BF01400372} {\bibfield  {journal} {\bibinfo  {journal} {Z. Phys.}\ }\textbf {\bibinfo {volume} {40}},\ \bibinfo {pages} {322} (\bibinfo {year} {1927})}\BibitemShut {NoStop}%
\bibitem [{\citenamefont {Sorokin}(2001)}]{Sorokin2001_Madelung}%
  \BibitemOpen
  \bibfield  {author} {\bibinfo {author} {\bibfnamefont {A.~L.}\ \bibnamefont {Sorokin}},\ }\bibfield  {title} {\bibinfo {title} {{\color{Black}{Madelung transformation for vortex flows of a perfect liquid}}},\ }\href {https://doi.org/10.1134/1.1401227} {\bibfield  {journal} {\bibinfo  {journal} {Dokl. Phys.}\ }\textbf {\bibinfo {volume} {46}},\ \bibinfo {pages} {576} (\bibinfo {year} {2001})}\BibitemShut {NoStop}%
\bibitem [{\citenamefont {Chern}\ \emph {et~al.}(2016)\citenamefont {Chern}, \citenamefont {Knöppel}, \citenamefont {Pinkall}, \citenamefont {Schröder},\ and\ \citenamefont {Weißmann}}]{Chern2016_Schrodinger}%
  \BibitemOpen
  \bibfield  {author} {\bibinfo {author} {\bibfnamefont {A.}~\bibnamefont {Chern}}, \bibinfo {author} {\bibfnamefont {F.}~\bibnamefont {Knöppel}}, \bibinfo {author} {\bibfnamefont {U.}~\bibnamefont {Pinkall}}, \bibinfo {author} {\bibfnamefont {P.}~\bibnamefont {Schröder}},\ and\ \bibinfo {author} {\bibfnamefont {S.}~\bibnamefont {Weißmann}},\ }\bibfield  {title} {\bibinfo {title} {{\color{Black}{Schrödinger's smoke}}},\ }\href {https://doi.org/10.1145/2897824.2925868} {\bibfield  {journal} {\bibinfo  {journal} {ACM Trans. Graph.}\ }\textbf {\bibinfo {volume} {35}},\ \bibinfo {pages} {1} (\bibinfo {year} {2016})}\BibitemShut {NoStop}%
\bibitem [{\citenamefont {Chern}\ \emph {et~al.}(2017)\citenamefont {Chern}, \citenamefont {Knöppel}, \citenamefont {Pinkall},\ and\ \citenamefont {Schröder}}]{Chern2017_Inside}%
  \BibitemOpen
  \bibfield  {author} {\bibinfo {author} {\bibfnamefont {A.}~\bibnamefont {Chern}}, \bibinfo {author} {\bibfnamefont {F.}~\bibnamefont {Knöppel}}, \bibinfo {author} {\bibfnamefont {U.}~\bibnamefont {Pinkall}},\ and\ \bibinfo {author} {\bibfnamefont {P.}~\bibnamefont {Schröder}},\ }\bibfield  {title} {\bibinfo {title} {{\color{Black}{Inside fluids: Clebsch maps for visualization and processing}}},\ }\href {https://doi.org/10.1145/3072959.3073591} {\bibfield  {journal} {\bibinfo  {journal} {ACM Trans. Graph.}\ }\textbf {\bibinfo {volume} {36}},\ \bibinfo {pages} {1} (\bibinfo {year} {2017})}\BibitemShut {NoStop}%
\bibitem [{\citenamefont {Meng}\ and\ \citenamefont {Yang}(2023)}]{Meng2023_Quantum}%
  \BibitemOpen
  \bibfield  {author} {\bibinfo {author} {\bibfnamefont {Z.}~\bibnamefont {Meng}}\ and\ \bibinfo {author} {\bibfnamefont {Y.}~\bibnamefont {Yang}},\ }\bibfield  {title} {\bibinfo {title} {{\color{Black}{Quantum computing of fluid dynamics using the hydrodynamic Schrödinger equation}}},\ }\href {https://doi.org/10.1103/PhysRevResearch.5.033182} {\bibfield  {journal} {\bibinfo  {journal} {Phys. Rev. Res.}\ }\textbf {\bibinfo {volume} {5}},\ \bibinfo {pages} {033182} (\bibinfo {year} {2023})}\BibitemShut {NoStop}%
\bibitem [{\citenamefont {Salasnich}\ \emph {et~al.}(2024)\citenamefont {Salasnich}, \citenamefont {Succi},\ and\ \citenamefont {Tiribocchi}}]{Salasnich2024_Quantum}%
  \BibitemOpen
  \bibfield  {author} {\bibinfo {author} {\bibfnamefont {L.}~\bibnamefont {Salasnich}}, \bibinfo {author} {\bibfnamefont {S.}~\bibnamefont {Succi}},\ and\ \bibinfo {author} {\bibfnamefont {A.}~\bibnamefont {Tiribocchi}},\ }\bibfield  {title} {\bibinfo {title} {{\color{Black}Quantum wave representation of dissipative fluids}},\ }\href {https://doi.org/10.1142/S0129183124501006} {\bibfield  {journal} {\bibinfo  {journal} {Int. J. Mod. Phys. C}\ ,\ \bibinfo {pages} {2450100}} (\bibinfo {year} {2024})}\BibitemShut {NoStop}%
\bibitem [{\citenamefont {Morrison}(1998)}]{Morrison1998_Hamiltonian}%
  \BibitemOpen
  \bibfield  {author} {\bibinfo {author} {\bibfnamefont {P.~J.}\ \bibnamefont {Morrison}},\ }\bibfield  {title} {\bibinfo {title} {{\color{Black}{Hamiltonian description of the ideal fluid}}},\ }\href {https://doi.org/10.1103/RevModPhys.70.467} {\bibfield  {journal} {\bibinfo  {journal} {Rev. Mod. Phys.}\ }\textbf {\bibinfo {volume} {70}},\ \bibinfo {pages} {467} (\bibinfo {year} {1998})}\BibitemShut {NoStop}%
\bibitem [{\citenamefont {Bohm}(1952)}]{Bohm1952_A}%
  \BibitemOpen
  \bibfield  {author} {\bibinfo {author} {\bibfnamefont {D.}~\bibnamefont {Bohm}},\ }\bibfield  {title} {\bibinfo {title} {{\color{Black}{A Suggested Interpretation of the Quantum Theory in Terms of "Hidden" Variables. I}}},\ }\href {https://doi.org/10.1103/PhysRev.85.166} {\bibfield  {journal} {\bibinfo  {journal} {Phys. Rev.}\ }\textbf {\bibinfo {volume} {85}},\ \bibinfo {pages} {166} (\bibinfo {year} {1952})}\BibitemShut {NoStop}%
\bibitem [{\citenamefont {Takabayasi}(1952)}]{Takabayasi1952_On}%
  \BibitemOpen
  \bibfield  {author} {\bibinfo {author} {\bibfnamefont {T.}~\bibnamefont {Takabayasi}},\ }\bibfield  {title} {\bibinfo {title} {{\color{Black}{On the formulation of quantum mechanics associated with classical pictures}}},\ }\href {https://doi.org/10.1143/ptp/8.2.143} {\bibfield  {journal} {\bibinfo  {journal} {Prog. Theor. Phys.}\ }\textbf {\bibinfo {volume} {8}},\ \bibinfo {pages} {143} (\bibinfo {year} {1952})}\BibitemShut {NoStop}%
\bibitem [{\citenamefont {Takabayasi}(1953)}]{Takabayasi1953_Remarks}%
  \BibitemOpen
  \bibfield  {author} {\bibinfo {author} {\bibfnamefont {T.}~\bibnamefont {Takabayasi}},\ }\bibfield  {title} {\bibinfo {title} {{\color{Black}{Remarks on the Formulation of Quantum Mechanics with Classical Pictures and on Relations between Linear Scalar Fields and Hydrodynamical Fields}}},\ }\href {https://doi.org/10.1143/ptp/9.3.187} {\bibfield  {journal} {\bibinfo  {journal} {Prog. Theor. Phys.}\ }\textbf {\bibinfo {volume} {9}},\ \bibinfo {pages} {187} (\bibinfo {year} {1953})}\BibitemShut {NoStop}%
\bibitem [{\citenamefont {Wyatt}(2005)}]{Wyatt2005_Quantum}%
  \BibitemOpen
  \bibfield  {author} {\bibinfo {author} {\bibfnamefont {R.~E.}\ \bibnamefont {Wyatt}},\ }\href@noop {} {\emph {\bibinfo {title} {{Quantum Dynamics with Trajectories -- Introduction to Quantum Hydrodynamics}}}},\ Vol.~\bibinfo {volume} {28}\ (\bibinfo  {publisher} {Springer},\ \bibinfo {year} {2005})\BibitemShut {NoStop}%
\bibitem [{\citenamefont {Bohm}(2012)}]{Bohm2012_Quantum}%
  \BibitemOpen
  \bibfield  {author} {\bibinfo {author} {\bibfnamefont {D.}~\bibnamefont {Bohm}},\ }\href@noop {} {\emph {\bibinfo {title} {{Quantum Theory}}}}\ (\bibinfo  {publisher} {Courier Corporation},\ \bibinfo {year} {2012})\BibitemShut {NoStop}%
\bibitem [{\citenamefont {Lu}\ and\ \citenamefont {Yang}(2023)}]{Lu2023_Quantum}%
  \BibitemOpen
  \bibfield  {author} {\bibinfo {author} {\bibfnamefont {Z.}~\bibnamefont {Lu}}\ and\ \bibinfo {author} {\bibfnamefont {Y.}~\bibnamefont {Yang}},\ }\href@noop {} {\bibinfo {title} {{\color{Black}{Quantum computing of reacting flows via Hamiltonian simulation}}}} (\bibinfo {year} {2023}),\ \Eprint {https://arxiv.org/abs/arXiv:2312.07893} {arXiv:2312.07893} \BibitemShut {NoStop}%
\bibitem [{\citenamefont {Meng}\ \emph {et~al.}(2024)\citenamefont {Meng}, \citenamefont {Zhong}, \citenamefont {Xu}, \citenamefont {Wang}, \citenamefont {Chen}, \citenamefont {Jin}, \citenamefont {Zhu}, \citenamefont {Gao}, \citenamefont {Wu}, \citenamefont {Zhang}, \citenamefont {Wang}, \citenamefont {Zou}, \citenamefont {Zhang}, \citenamefont {Cui}, \citenamefont {Shen}, \citenamefont {Bao}, \citenamefont {Zhu}, \citenamefont {Tan}, \citenamefont {Li}, \citenamefont {Zhang}, \citenamefont {Xiong}, \citenamefont {Li}, \citenamefont {Guo}, \citenamefont {Wang}, \citenamefont {Song}, \citenamefont {Wang},\ and\ \citenamefont {Yang}}]{Meng2024_Simulating}%
  \BibitemOpen
  \bibfield  {author} {\bibinfo {author} {\bibfnamefont {Z.}~\bibnamefont {Meng}}, \bibinfo {author} {\bibfnamefont {J.}~\bibnamefont {Zhong}}, \bibinfo {author} {\bibfnamefont {S.}~\bibnamefont {Xu}}, \bibinfo {author} {\bibfnamefont {K.}~\bibnamefont {Wang}}, \bibinfo {author} {\bibfnamefont {J.}~\bibnamefont {Chen}}, \bibinfo {author} {\bibfnamefont {F.}~\bibnamefont {Jin}}, \bibinfo {author} {\bibfnamefont {X.}~\bibnamefont {Zhu}}, \bibinfo {author} {\bibfnamefont {Y.}~\bibnamefont {Gao}}, \bibinfo {author} {\bibfnamefont {Y.}~\bibnamefont {Wu}}, \bibinfo {author} {\bibfnamefont {C.}~\bibnamefont {Zhang}}, \emph {et~al.},\ }\href {https://doi.org/10.48550/arXiv.2404.15878} {\bibinfo {title} {{\color{Black}{Simulating unsteady fluid flows on a superconducting quantum processor}}}} (\bibinfo {year} {2024}),\ \Eprint {https://arxiv.org/abs/arXiv:2404.15878} {arXiv:2404.15878} \BibitemShut {NoStop}%
\bibitem [{\citenamefont {Feynman}(1982)}]{Feynman1982_Simulating}%
  \BibitemOpen
  \bibfield  {author} {\bibinfo {author} {\bibfnamefont {R.~P.}\ \bibnamefont {Feynman}},\ }\bibfield  {title} {\bibinfo {title} {{\color{Black}Simulating physics with computers}},\ }\href {https://doi.org/10.1007/BF02650179} {\bibfield  {journal} {\bibinfo  {journal} {Int. J. Theor. Phys.}\ }\textbf {\bibinfo {volume} {21}},\ \bibinfo {pages} {467} (\bibinfo {year} {1982})}\BibitemShut {NoStop}%
\bibitem [{\citenamefont {Givi}\ \emph {et~al.}(2020)\citenamefont {Givi}, \citenamefont {Daley}, \citenamefont {Mavriplis},\ and\ \citenamefont {Malik}}]{Givi2020_Quantum}%
  \BibitemOpen
  \bibfield  {author} {\bibinfo {author} {\bibfnamefont {P.}~\bibnamefont {Givi}}, \bibinfo {author} {\bibfnamefont {A.~J.}\ \bibnamefont {Daley}}, \bibinfo {author} {\bibfnamefont {D.}~\bibnamefont {Mavriplis}},\ and\ \bibinfo {author} {\bibfnamefont {M.}~\bibnamefont {Malik}},\ }\bibfield  {title} {\bibinfo {title} {{\color{Black}{Quantum speedup for aeroscience and engineering}}},\ }\href {https://doi.org/10.2514/1.J059183} {\bibfield  {journal} {\bibinfo  {journal} {AIAA J.}\ }\textbf {\bibinfo {volume} {58}},\ \bibinfo {pages} {8} (\bibinfo {year} {2020})}\BibitemShut {NoStop}%
\bibitem [{\citenamefont {Succi}\ \emph {et~al.}(2023)\citenamefont {Succi}, \citenamefont {Itani}, \citenamefont {Sreenivasan},\ and\ \citenamefont {Steijl}}]{Succi2023_Quantum}%
  \BibitemOpen
  \bibfield  {author} {\bibinfo {author} {\bibfnamefont {S.}~\bibnamefont {Succi}}, \bibinfo {author} {\bibfnamefont {W.}~\bibnamefont {Itani}}, \bibinfo {author} {\bibfnamefont {K.}~\bibnamefont {Sreenivasan}},\ and\ \bibinfo {author} {\bibfnamefont {R.}~\bibnamefont {Steijl}},\ }\bibfield  {title} {\bibinfo {title} {{\color{Black}{Quantum computing for fluids: Where do we stand?}}},\ }\href {https://doi.org/10.1209/0295-5075/acfdc7} {\bibfield  {journal} {\bibinfo  {journal} {Europhys. Lett.}\ }\textbf {\bibinfo {volume} {144}},\ \bibinfo {pages} {10001} (\bibinfo {year} {2023})}\BibitemShut {NoStop}%
\bibitem [{\citenamefont {Bharadwaj}\ and\ \citenamefont {Sreenivasan}(2023)}]{Bharadwaj2023_Hybrid}%
  \BibitemOpen
  \bibfield  {author} {\bibinfo {author} {\bibfnamefont {S.~S.}\ \bibnamefont {Bharadwaj}}\ and\ \bibinfo {author} {\bibfnamefont {K.~R.}\ \bibnamefont {Sreenivasan}},\ }\bibfield  {title} {\bibinfo {title} {{\color{Black}{Hybrid quantum algorithms for flow problems}}},\ }\href {https://doi.org/10.1073/pnas.2311014120} {\bibfield  {journal} {\bibinfo  {journal} {Proc. Natl. Acad. Sci. U.S.A.}\ }\textbf {\bibinfo {volume} {120}},\ \bibinfo {pages} {e2311014120} (\bibinfo {year} {2023})}\BibitemShut {NoStop}%
\bibitem [{\citenamefont {Meng}\ and\ \citenamefont {Yang}(2024)}]{Meng2024_Lagrangian}%
  \BibitemOpen
  \bibfield  {author} {\bibinfo {author} {\bibfnamefont {Z.}~\bibnamefont {Meng}}\ and\ \bibinfo {author} {\bibfnamefont {Y.}~\bibnamefont {Yang}},\ }\bibfield  {title} {\bibinfo {title} {{\color{Black}{Lagrangian dynamics and regularity of the spin Euler equation}}},\ }\href {https://doi.org/10.1017/jfm.2024.319} {\bibfield  {journal} {\bibinfo  {journal} {J. Fluid Mech.}\ }\textbf {\bibinfo {volume} {985}},\ \bibinfo {pages} {A34} (\bibinfo {year} {2024})}\BibitemShut {NoStop}%
\bibitem [{\citenamefont {Xiong}\ \emph {et~al.}(2022)\citenamefont {Xiong}, \citenamefont {Wang}, \citenamefont {Wang},\ and\ \citenamefont {Zhu}}]{Xiong2022Clebsch}%
  \BibitemOpen
  \bibfield  {author} {\bibinfo {author} {\bibfnamefont {S.}~\bibnamefont {Xiong}}, \bibinfo {author} {\bibfnamefont {Z.}~\bibnamefont {Wang}}, \bibinfo {author} {\bibfnamefont {M.}~\bibnamefont {Wang}},\ and\ \bibinfo {author} {\bibfnamefont {B.}~\bibnamefont {Zhu}},\ }\bibfield  {title} {\bibinfo {title} {{\color{Black}{A Clebsch method for free-surface vortical flow simulation}}},\ }\href@noop {} {\bibfield  {journal} {\bibinfo  {journal} {ACM Trans. Graph.}\ }\textbf {\bibinfo {volume} {41}} (\bibinfo {year} {2022})}\BibitemShut {NoStop}%
\bibitem [{\citenamefont {Pesci}\ \emph {et~al.}(2005)\citenamefont {Pesci}, \citenamefont {Goldstein},\ and\ \citenamefont {Uys}}]{Pesci2005_Mapping}%
  \BibitemOpen
  \bibfield  {author} {\bibinfo {author} {\bibfnamefont {A.~I.}\ \bibnamefont {Pesci}}, \bibinfo {author} {\bibfnamefont {R.~E.}\ \bibnamefont {Goldstein}},\ and\ \bibinfo {author} {\bibfnamefont {H.}~\bibnamefont {Uys}},\ }\bibfield  {title} {\bibinfo {title} {{\color{Black}{Mapping of the classical kinetic balance equations onto the Pauli equation}}},\ }\href {https://doi.org/10.1088/0951-7715/18/1/012} {\bibfield  {journal} {\bibinfo  {journal} {Nonlinearity}\ }\textbf {\bibinfo {volume} {18}},\ \bibinfo {pages} {227} (\bibinfo {year} {2005})}\BibitemShut {NoStop}%
\bibitem [{\citenamefont {Saffman}(1993)}]{Saffman1993_Vortex}%
  \BibitemOpen
  \bibfield  {author} {\bibinfo {author} {\bibfnamefont {P.~G.}\ \bibnamefont {Saffman}},\ }\href@noop {} {\emph {\bibinfo {title} {Vortex Dynamics}}},\ Cambridge Monographs on Mechanics\ (\bibinfo  {publisher} {Cambridge University Press},\ \bibinfo {year} {1993})\BibitemShut {NoStop}%
\bibitem [{\citenamefont {Mueller}\ and\ \citenamefont {Ho}(2002)}]{Mueller2002_Two}%
  \BibitemOpen
  \bibfield  {author} {\bibinfo {author} {\bibfnamefont {E.~J.}\ \bibnamefont {Mueller}}\ and\ \bibinfo {author} {\bibfnamefont {T.-L.}\ \bibnamefont {Ho}},\ }\bibfield  {title} {\bibinfo {title} {{\color{Black}{Two-component Bose-Einstein condensates with a large number of vortices}}},\ }\href {https://doi.org/10.1103/PhysRevLett.88.180403} {\bibfield  {journal} {\bibinfo  {journal} {Phys. Rev. Lett.}\ }\textbf {\bibinfo {volume} {88}},\ \bibinfo {pages} {180403} (\bibinfo {year} {2002})}\BibitemShut {NoStop}%
\bibitem [{\citenamefont {Kasamatsu}\ \emph {et~al.}(2003)\citenamefont {Kasamatsu}, \citenamefont {Tsubota},\ and\ \citenamefont {Ueda}}]{Kasamatsu2003_Vortex}%
  \BibitemOpen
  \bibfield  {author} {\bibinfo {author} {\bibfnamefont {K.}~\bibnamefont {Kasamatsu}}, \bibinfo {author} {\bibfnamefont {M.}~\bibnamefont {Tsubota}},\ and\ \bibinfo {author} {\bibfnamefont {M.}~\bibnamefont {Ueda}},\ }\bibfield  {title} {\bibinfo {title} {{\color{Black}{Vortex phase diagram in rotating two-component Bose-Einstein condensates}}},\ }\href {https://doi.org/10.1103/PhysRevLett.91.150406} {\bibfield  {journal} {\bibinfo  {journal} {Phys. Rev. Lett.}\ }\textbf {\bibinfo {volume} {91}},\ \bibinfo {pages} {150406} (\bibinfo {year} {2003})}\BibitemShut {NoStop}%
\bibitem [{\citenamefont {Liu}\ and\ \citenamefont {Kengne}(2019)}]{Liu2019_Schrodinger}%
  \BibitemOpen
  \bibfield  {author} {\bibinfo {author} {\bibfnamefont {W.-M.}\ \bibnamefont {Liu}}\ and\ \bibinfo {author} {\bibfnamefont {E.}~\bibnamefont {Kengne}},\ }\href {https://doi.org/10.1007/978-981-13-6581-2_1} {\emph {\bibinfo {title} {{Schr\"odinger Equations in Nonlinear Systems}}}}\ (\bibinfo  {publisher} {Springer Singapore},\ \bibinfo {year} {2019})\BibitemShut {NoStop}%
\bibitem [{\citenamefont {Greene}(1993)}]{Greene1993_Reconnection}%
  \BibitemOpen
  \bibfield  {author} {\bibinfo {author} {\bibfnamefont {J.~M.}\ \bibnamefont {Greene}},\ }\bibfield  {title} {\bibinfo {title} {{\color{Black}{Reconnection of vorticity lines and magnetic lines}}},\ }\href {https://doi.org/10.1063/1.860718} {\bibfield  {journal} {\bibinfo  {journal} {Phys. Fluids B}\ }\textbf {\bibinfo {volume} {5}},\ \bibinfo {pages} {2355} (\bibinfo {year} {1993})}\BibitemShut {NoStop}%
\bibitem [{\citenamefont {Huang}\ \emph {et~al.}(1997)\citenamefont {Huang}, \citenamefont {K\"{u}pper},\ and\ \citenamefont {Masbaum}}]{Huang1997_Computation}%
  \BibitemOpen
  \bibfield  {author} {\bibinfo {author} {\bibfnamefont {M.}~\bibnamefont {Huang}}, \bibinfo {author} {\bibfnamefont {T.}~\bibnamefont {K\"{u}pper}},\ and\ \bibinfo {author} {\bibfnamefont {N.}~\bibnamefont {Masbaum}},\ }\bibfield  {title} {\bibinfo {title} {{\color{Black}Computation of invariant tori by the Fourier methods}},\ }\href {https://doi.org/10.1137/S1064827593258826} {\bibfield  {journal} {\bibinfo  {journal} {SIAM J. Sci. Comput.}\ }\textbf {\bibinfo {volume} {18}},\ \bibinfo {pages} {918} (\bibinfo {year} {1997})}\BibitemShut {NoStop}%
\bibitem [{\citenamefont {Hao}\ \emph {et~al.}(2019)\citenamefont {Hao}, \citenamefont {Xiong},\ and\ \citenamefont {Yang}}]{Hao2019_Tracking}%
  \BibitemOpen
  \bibfield  {author} {\bibinfo {author} {\bibfnamefont {J.}~\bibnamefont {Hao}}, \bibinfo {author} {\bibfnamefont {S.}~\bibnamefont {Xiong}},\ and\ \bibinfo {author} {\bibfnamefont {Y.}~\bibnamefont {Yang}},\ }\bibfield  {title} {\bibinfo {title} {{\color{Black}{Tracking vortex surfaces frozen in the virtual velocity in non-ideal flows}}},\ }\href {https://doi.org/10.1017/jfm.2018.1014} {\bibfield  {journal} {\bibinfo  {journal} {J. Fluid Mech.}\ }\textbf {\bibinfo {volume} {863}},\ \bibinfo {pages} {513} (\bibinfo {year} {2019})}\BibitemShut {NoStop}%
\bibitem [{\citenamefont {Yang}\ \emph {et~al.}(2023)\citenamefont {Yang}, \citenamefont {Xiong},\ and\ \citenamefont {Lu}}]{Yang2023_Applications}%
  \BibitemOpen
  \bibfield  {author} {\bibinfo {author} {\bibfnamefont {Y.}~\bibnamefont {Yang}}, \bibinfo {author} {\bibfnamefont {S.}~\bibnamefont {Xiong}},\ and\ \bibinfo {author} {\bibfnamefont {Z.}~\bibnamefont {Lu}},\ }\bibfield  {title} {\bibinfo {title} {{\color{Black}{Applications of the vortex-surface field to flow visualization, modelling and simulation}}},\ }\href {https://doi.org/10.1017/flo.2023.27} {\bibfield  {journal} {\bibinfo  {journal} {Flow}\ }\textbf {\bibinfo {volume} {3}},\ \bibinfo {pages} {E33} (\bibinfo {year} {2023})}\BibitemShut {NoStop}%
\bibitem [{\citenamefont {Eyink}(2010)}]{Eyink2010_Stochastic}%
  \BibitemOpen
  \bibfield  {author} {\bibinfo {author} {\bibfnamefont {G.~L.}\ \bibnamefont {Eyink}},\ }\bibfield  {title} {\bibinfo {title} {{\color{Black}{Stochastic least-action principle for the incompressible Navier–Stokes equation}}},\ }\href {https://doi.org/10.1016/j.physd.2008.11.011} {\bibfield  {journal} {\bibinfo  {journal} {Physica D}\ }\textbf {\bibinfo {volume} {239}},\ \bibinfo {pages} {1236} (\bibinfo {year} {2010})}\BibitemShut {NoStop}%
\bibitem [{\citenamefont {Frisch}\ \emph {et~al.}(1986)\citenamefont {Frisch}, \citenamefont {Hasslacher},\ and\ \citenamefont {Pomeau}}]{Frisch1986_Lattice}%
  \BibitemOpen
  \bibfield  {author} {\bibinfo {author} {\bibfnamefont {U.}~\bibnamefont {Frisch}}, \bibinfo {author} {\bibfnamefont {B.}~\bibnamefont {Hasslacher}},\ and\ \bibinfo {author} {\bibfnamefont {Y.}~\bibnamefont {Pomeau}},\ }\bibfield  {title} {\bibinfo {title} {{\color{Black}{Lattice-gas automata for the Navier-Stokes equation}}},\ }\href {https://doi.org/10.1103/PhysRevLett.56.1505} {\bibfield  {journal} {\bibinfo  {journal} {Phys. Rev. Lett.}\ }\textbf {\bibinfo {volume} {56}},\ \bibinfo {pages} {1505} (\bibinfo {year} {1986})}\BibitemShut {NoStop}%
\bibitem [{\citenamefont {Yepez}(2001)}]{Yepez2001_Quantum}%
  \BibitemOpen
  \bibfield  {author} {\bibinfo {author} {\bibfnamefont {J.}~\bibnamefont {Yepez}},\ }\bibfield  {title} {\bibinfo {title} {{\color{Black}{Quantum lattice-gas model for computational fluid dynamics}}},\ }\href {https://doi.org/10.1103/PhysRevE.63.046702} {\bibfield  {journal} {\bibinfo  {journal} {Phys. Rev. E}\ }\textbf {\bibinfo {volume} {63}},\ \bibinfo {pages} {046702} (\bibinfo {year} {2001})}\BibitemShut {NoStop}%
\bibitem [{\citenamefont {Bransden}\ and\ \citenamefont {Joachain}(2003)}]{Bransden2003_Physics}%
  \BibitemOpen
  \bibfield  {author} {\bibinfo {author} {\bibfnamefont {B.~H.}\ \bibnamefont {Bransden}}\ and\ \bibinfo {author} {\bibfnamefont {C.~J.}\ \bibnamefont {Joachain}},\ }\href {https://catalogue.library.cern/literature/m0c8r-z7653} {\emph {\bibinfo {title} {{Physics of Atoms and Molecules}}}}\ (\bibinfo  {publisher} {Pearson Education India},\ \bibinfo {year} {2003})\BibitemShut {NoStop}%
\bibitem [{\citenamefont {Ashida}\ \emph {et~al.}(2021)\citenamefont {Ashida}, \citenamefont {Gong},\ and\ \citenamefont {Ueda}}]{Ashida2021_Non}%
  \BibitemOpen
  \bibfield  {author} {\bibinfo {author} {\bibfnamefont {Y.}~\bibnamefont {Ashida}}, \bibinfo {author} {\bibfnamefont {Z.}~\bibnamefont {Gong}},\ and\ \bibinfo {author} {\bibfnamefont {M.}~\bibnamefont {Ueda}},\ }\bibfield  {title} {\bibinfo {title} {{\color{Black}{Non-Hermitian physics}}},\ }\href {https://doi.org/10.1080/00018732.2021.1876991} {\bibfield  {journal} {\bibinfo  {journal} {Adv. Phys.}\ }\textbf {\bibinfo {volume} {69}},\ \bibinfo {pages} {249} (\bibinfo {year} {2021})}\BibitemShut {NoStop}%
\bibitem [{\citenamefont {Suzuki}(1990)}]{Suzuki1990_Fractal}%
  \BibitemOpen
  \bibfield  {author} {\bibinfo {author} {\bibfnamefont {M.}~\bibnamefont {Suzuki}},\ }\bibfield  {title} {\bibinfo {title} {{\color{Black}{Fractal decomposition of exponential operators with applications to many-body theories and Monte Carlo simulations}}},\ }\href {https://doi.org/https://doi.org/10.1016/0375-9601(90)90962-N} {\bibfield  {journal} {\bibinfo  {journal} {Phys. Lett. A}\ }\textbf {\bibinfo {volume} {146}},\ \bibinfo {pages} {319} (\bibinfo {year} {1990})}\BibitemShut {NoStop}%
\bibitem [{\citenamefont {Suzuki}(1991)}]{Suzuki1991_General}%
  \BibitemOpen
  \bibfield  {author} {\bibinfo {author} {\bibfnamefont {M.}~\bibnamefont {Suzuki}},\ }\bibfield  {title} {\bibinfo {title} {{\color{Black}{General theory of fractal path integrals with applications to many-body theories and statistical physics}}},\ }\href {https://doi.org/10.1063/1.529425} {\bibfield  {journal} {\bibinfo  {journal} {J. Math. Phys.}\ }\textbf {\bibinfo {volume} {32}},\ \bibinfo {pages} {400} (\bibinfo {year} {1991})}\BibitemShut {NoStop}%
\bibitem [{cod()}]{code}%
  \BibitemOpen
  \href@noop {} {\bibinfo {title} {{\color{Black}The code is available at https://github.com/YYgroup/SPE.}}}\BibitemShut {Stop}%
\bibitem [{\citenamefont {Yang}\ \emph {et~al.}(2021)\citenamefont {Yang}, \citenamefont {Xiong}, \citenamefont {Zhang}, \citenamefont {Feng}, \citenamefont {Liu},\ and\ \citenamefont {Zhu}}]{Yang2021_Clebsch}%
  \BibitemOpen
  \bibfield  {author} {\bibinfo {author} {\bibfnamefont {S.}~\bibnamefont {Yang}}, \bibinfo {author} {\bibfnamefont {S.}~\bibnamefont {Xiong}}, \bibinfo {author} {\bibfnamefont {Y.}~\bibnamefont {Zhang}}, \bibinfo {author} {\bibfnamefont {F.}~\bibnamefont {Feng}}, \bibinfo {author} {\bibfnamefont {J.}~\bibnamefont {Liu}},\ and\ \bibinfo {author} {\bibfnamefont {B.}~\bibnamefont {Zhu}},\ }\bibfield  {title} {\bibinfo {title} {{\color{Black}{Clebsch gauge fluid}}},\ }\href {https://doi.org/10.1145/3450626.3459866} {\bibfield  {journal} {\bibinfo  {journal} {ACM Trans. Graphics}\ }\textbf {\bibinfo {volume} {40}},\ \bibinfo {pages} {1} (\bibinfo {year} {2021})}\BibitemShut {NoStop}%
\bibitem [{\citenamefont {Taylor}\ and\ \citenamefont {Green}(1937)}]{Taylor1937_Mechanism}%
  \BibitemOpen
  \bibfield  {author} {\bibinfo {author} {\bibfnamefont {G.~I.}\ \bibnamefont {Taylor}}\ and\ \bibinfo {author} {\bibfnamefont {A.~E.}\ \bibnamefont {Green}},\ }\bibfield  {title} {\bibinfo {title} {{\color{Black}Mechanism of the production of small eddies from large ones}},\ }\href {https://doi.org/10.1098/rspa.1937.0036} {\bibfield  {journal} {\bibinfo  {journal} {Proc. R. Soc. London Ser. A-Math. Phys. Eng. Sci.}\ }\textbf {\bibinfo {volume} {158}},\ \bibinfo {pages} {499} (\bibinfo {year} {1937})}\BibitemShut {NoStop}%
\bibitem [{\citenamefont {Su}\ \emph {et~al.}(2024)\citenamefont {Su}, \citenamefont {Xiong},\ and\ \citenamefont {Yang}}]{Su2024_Quantum}%
  \BibitemOpen
  \bibfield  {author} {\bibinfo {author} {\bibfnamefont {H.}~\bibnamefont {Su}}, \bibinfo {author} {\bibfnamefont {S.}~\bibnamefont {Xiong}},\ and\ \bibinfo {author} {\bibfnamefont {Y.}~\bibnamefont {Yang}},\ }\href {https://doi.org/10.48550/arXiv.2406.04652} {\bibinfo {title} {{\color{Black}{Quantum state preparation for a velocity field based on the spherical Clebsch wave function}}}} (\bibinfo {year} {2024}),\ \Eprint {https://arxiv.org/abs/arXiv:2406.04652} {arXiv:2406.04652} \BibitemShut {NoStop}%
\bibitem [{\citenamefont {Lloyd}\ \emph {et~al.}(2020)\citenamefont {Lloyd}, \citenamefont {Palma}, \citenamefont {Gokler}, \citenamefont {Kiani}, \citenamefont {Liu}, \citenamefont {Marvian}, \citenamefont {Tennie},\ and\ \citenamefont {Palmer}}]{Lloyd2020_Quantum}%
  \BibitemOpen
  \bibfield  {author} {\bibinfo {author} {\bibfnamefont {S.}~\bibnamefont {Lloyd}}, \bibinfo {author} {\bibfnamefont {G.~D.}\ \bibnamefont {Palma}}, \bibinfo {author} {\bibfnamefont {C.}~\bibnamefont {Gokler}}, \bibinfo {author} {\bibfnamefont {B.}~\bibnamefont {Kiani}}, \bibinfo {author} {\bibfnamefont {Z.-W.}\ \bibnamefont {Liu}}, \bibinfo {author} {\bibfnamefont {M.}~\bibnamefont {Marvian}}, \bibinfo {author} {\bibfnamefont {F.}~\bibnamefont {Tennie}},\ and\ \bibinfo {author} {\bibfnamefont {T.}~\bibnamefont {Palmer}},\ }\href@noop {} {\bibinfo {title} {{\color{Black}{Quantum algorithm for nonlinear differential equations}}}} (\bibinfo {year} {2020}),\ \Eprint {https://arxiv.org/abs/arXiv:2011.06571} {arXiv:2011.06571} \BibitemShut {NoStop}%
\bibitem [{\citenamefont {Großardt}(2024)}]{Großardt2024_Nonlinear}%
  \BibitemOpen
  \bibfield  {author} {\bibinfo {author} {\bibfnamefont {A.}~\bibnamefont {Großardt}},\ }\href {https://doi.org/10.48550/arXiv.2403.10102} {\bibinfo {title} {{\color{Black}{Nonlinear-ancilla aided quantum algorithm for nonlinear Schrödinger equations}}}} (\bibinfo {year} {2024}),\ \Eprint {https://arxiv.org/abs/arXiv:2403.10102} {arXiv:2403.10102} \BibitemShut {NoStop}%
\bibitem [{\citenamefont {Brüstle}\ and\ \citenamefont {Wiebe}(2024)}]{Brüstle2024_Quantum}%
  \BibitemOpen
  \bibfield  {author} {\bibinfo {author} {\bibfnamefont {N.}~\bibnamefont {Brüstle}}\ and\ \bibinfo {author} {\bibfnamefont {N.}~\bibnamefont {Wiebe}},\ }\href {https://doi.org/10.48550/arXiv.2407.07685} {\bibinfo {title} {{\color{Black}{Quantum and classical algorithms for nonlinear unitary dynamics}}}} (\bibinfo {year} {2024}),\ \Eprint {https://arxiv.org/abs/arXiv:2407.07685} {arXiv:2407.07685} \BibitemShut {NoStop}%
\bibitem [{\citenamefont {Jin}\ \emph {et~al.}(2023)\citenamefont {Jin}, \citenamefont {Liu},\ and\ \citenamefont {Yu}}]{Jin2023_Quantum}%
  \BibitemOpen
  \bibfield  {author} {\bibinfo {author} {\bibfnamefont {S.}~\bibnamefont {Jin}}, \bibinfo {author} {\bibfnamefont {N.}~\bibnamefont {Liu}},\ and\ \bibinfo {author} {\bibfnamefont {Y.}~\bibnamefont {Yu}},\ }\bibfield  {title} {\bibinfo {title} {{\color{Black}{Quantum simulation of partial differential equations: Applications and detailed analysis}}},\ }\href {https://doi.org/10.1103/PhysRevA.108.032603} {\bibfield  {journal} {\bibinfo  {journal} {Phys. Rev. A}\ }\textbf {\bibinfo {volume} {108}},\ \bibinfo {pages} {032603} (\bibinfo {year} {2023})}\BibitemShut {NoStop}%
\bibitem [{\citenamefont {Müller}\ \emph {et~al.}(2021)\citenamefont {Müller}, \citenamefont {Polanco},\ and\ \citenamefont {Krstulovic}}]{Muller2021_Intermittency}%
  \BibitemOpen
  \bibfield  {author} {\bibinfo {author} {\bibfnamefont {N.~P.}\ \bibnamefont {Müller}}, \bibinfo {author} {\bibfnamefont {J.~I.}\ \bibnamefont {Polanco}},\ and\ \bibinfo {author} {\bibfnamefont {G.}~\bibnamefont {Krstulovic}},\ }\bibfield  {title} {\bibinfo {title} {{\color{Black}{Intermittency of velocity circulation in quantum turbulence}}},\ }\href {https://doi.org/10.1103/PhysRevX.11.011053} {\bibfield  {journal} {\bibinfo  {journal} {Phys. Rev. X}\ }\textbf {\bibinfo {volume} {11}},\ \bibinfo {pages} {011053} (\bibinfo {year} {2021})}\BibitemShut {NoStop}%
\bibitem [{\citenamefont {Shen}\ \emph {et~al.}(2024)\citenamefont {Shen}, \citenamefont {Yao},\ and\ \citenamefont {Yang}}]{Shen2024_Weaving}%
  \BibitemOpen
  \bibfield  {author} {\bibinfo {author} {\bibfnamefont {W.}~\bibnamefont {Shen}}, \bibinfo {author} {\bibfnamefont {J.}~\bibnamefont {Yao}},\ and\ \bibinfo {author} {\bibfnamefont {Y.}~\bibnamefont {Yang}},\ }\href@noop {} {\bibinfo {title} {{\color{Black}{Designing turbulence with entangled vortices}}}} (\bibinfo {year} {2024}),\ \bibinfo {note} {{Proc. Natl. Acad. Sci. U.S.A. (in press)}},\ \Eprint {https://arxiv.org/abs/arXiv:2401.11149} {arXiv:2401.11149} \BibitemShut {NoStop}%
\bibitem [{\citenamefont {Bergholtz}\ \emph {et~al.}(2021)\citenamefont {Bergholtz}, \citenamefont {Budich},\ and\ \citenamefont {Kunst}}]{Bergholtz2021_Exceptional}%
  \BibitemOpen
  \bibfield  {author} {\bibinfo {author} {\bibfnamefont {E.~J.}\ \bibnamefont {Bergholtz}}, \bibinfo {author} {\bibfnamefont {J.~C.}\ \bibnamefont {Budich}},\ and\ \bibinfo {author} {\bibfnamefont {F.~K.}\ \bibnamefont {Kunst}},\ }\bibfield  {title} {\bibinfo {title} {{\color{Black}{Exceptional topology of non-Hermitian systems}}},\ }\href {https://doi.org/10.1103/RevModPhys.93.015005} {\bibfield  {journal} {\bibinfo  {journal} {Rev. Mod. Phys.}\ }\textbf {\bibinfo {volume} {93}},\ \bibinfo {pages} {015005} (\bibinfo {year} {2021})}\BibitemShut {NoStop}%
\bibitem [{\citenamefont {Ding}\ \emph {et~al.}(2022)\citenamefont {Ding}, \citenamefont {Fang},\ and\ \citenamefont {Ma}}]{Ding2022_Non}%
  \BibitemOpen
  \bibfield  {author} {\bibinfo {author} {\bibfnamefont {K.}~\bibnamefont {Ding}}, \bibinfo {author} {\bibfnamefont {C.}~\bibnamefont {Fang}},\ and\ \bibinfo {author} {\bibfnamefont {G.}~\bibnamefont {Ma}},\ }\bibfield  {title} {\bibinfo {title} {{\color{Black}{Non-Hermitian topology and exceptional-point geometries}}},\ }\href {https://doi.org/10.1038/s42254-022-00516-5} {\bibfield  {journal} {\bibinfo  {journal} {Nat. Rev. Phys.}\ }\textbf {\bibinfo {volume} {4}},\ \bibinfo {pages} {745} (\bibinfo {year} {2022})}\BibitemShut {NoStop}%
\end{thebibliography}
%

\let\addcontentsline\oldaddcontentsline

\end{document}